\documentclass[a4paper,10pt]{article}
\usepackage[utf8]{inputenc}
\usepackage{authblk}
\usepackage{chngcntr}
\usepackage{graphicx}
\usepackage{subfig}
\usepackage{textcomp}
\usepackage{booktabs}
\usepackage{ulem}
\usepackage{amsmath}
\usepackage{amssymb}
\usepackage{color}
\usepackage{framed}
\usepackage{txfonts}
\usepackage{upgreek}
\usepackage{times}
\usepackage{bm}
\usepackage{mathbbol}
\usepackage{tensor}
\usepackage{empheq}
\usepackage[makeroom]{cancel}
\usepackage{needspace}
\usepackage{xspace}
\usepackage{relsize}
\usepackage{verbatim}
\usepackage{float}
\usepackage{bigints}
\usepackage{xcolor}
\begin{document}
\definecolor{shadecolor}{rgb}{0.93,0.93,0.93}
\definecolor{darkred}{rgb}{0.64,0.34,0.37}
\definecolor{dunkelgrau}{rgb}{0.33,0.33,0.33}
\definecolor{eqemphcolor}{rgb}{1,1,1} 
\definecolor{eqemphcolor}{rgb}{0.93,0.93,0.93} 
\definecolor{eqemphcolor}{rgb}{1,1,0.75} 
\empheqset{box={\fboxsep=2pt\colorbox{eqemphcolor}}}
\bibliographystyle{plain}    

%
%
%
\newcommand{\RB}{\mathbb{R}}
\newcommand{\MB}{\mathbb{M}}
\newcommand{\NB}{\mathbb{N}}
\newcommand{\CB}{\mathbb{C}}
\renewcommand{\MB}{\mathbb{M}}
\newcommand{\AB}{\mathbb{A}}
\newcommand{\BB}{\mathbb{B}}
\newcommand{\im}{\mathrm{i}}
\newcommand{\FC}{F}
\newcommand{\e}{\mathrm{e}}
\newcommand{\Lag}{L}
\newcommand{\Lae}{\Lag_{\e}}
\newcommand{\HCv}{E}
\newcommand{\Hf}{H}
\newcommand{\He}{\Hf_{\e}}
\newcommand{\HCp}{\Hf^{\prime}}
\newcommand{\HCe}{\He^{\prime}}
\newcommand{\Hv}{e}
\newcommand{\HD}{\mathcal{\Hf}}
\newcommand{\HCC}{\Hf^{\prime\prime}}
\newcommand{\HCd}{\mathcal{\Hf}}
\newcommand{\FCd}{\mathcal{\FC}}
\newcommand{\LCd}{\mathcal{\Lag}}
\newcommand{\LTd}{\mathcal{T}}
\newcommand{\KCd}{\mathcal{K}}
\newcommand{\QCd}{\mathcal{Q}}
\newcommand{\qCd}{\mathcal{q}}
\newcommand{\Dderr}{\overrightarrow{\mathcal{D}}}
\newcommand{\Dderl}{\overleftarrow{\mathcal{D}}}
\newcommand{\bx}{\pmb{x}}
\newcommand{\by}{\pmb{y}}
\newcommand{\bk}{\pmb{k}}
\newcommand{\be}{\pmb{e}}
\newcommand{\bef}{\pmb{f}}
\newcommand{\bg}{\pmb{g}}
\newcommand{\bp}{\pmb{p}}
\newcommand{\bq}{\pmb{q}}
\newcommand{\ba}{\pmb{a}}
\newcommand{\bA}{\pmb{A}}
\newcommand{\bb}{\pmb{b}}
\newcommand{\bB}{\pmb{B}}
\newcommand{\bF}{\FC^{\mu}}
\newcommand{\bP}{\pmb{P}}
\newcommand{\bQ}{\pmb{Q}}
\newcommand{\bPhi}{\pmb{\Phi}}
\newcommand{\bphi}{\pmb{\varphi}}
\newcommand{\bPsi}{\pmb{\Psi}}
\newcommand{\bpsi}{\pmb{\psi}}
\newcommand{\bPi}{\pmb{\Pi}}
\newcommand{\bpi}{\pmb{\pi}}
\newcommand{\bvarphi}{\pmb{\varphi}}
\newcommand{\rmi}{\im}
\newcommand{\CO}{\mathcal{O}}
\renewcommand{\d}{\,\mathrm{d}}
\newcommand{\ds}{\d{s}}
\newcommand{\dt}{\d{t}}
\newcommand{\onehalf}{{\textstyle\frac{1}{2}}}
\newcommand{\twothird}{{\textstyle\frac{2}{3}}}
\newcommand{\onethird}{{\textstyle\frac{1}{3}}}
\newcommand{\quarter}{{\textstyle\frac{1}{4}}}
\newcommand{\threequarter}{{\textstyle\frac{3}{4}}}
\newcommand{\fourthird}{{\textstyle\frac{4}{3}}}
\newcommand{\oneeights}{{\textstyle\frac{1}{8}}}
\newcommand{\ihalf}{{\textstyle\frac{\rmi}{2}}}
\newcommand{\iquarter}{{\textstyle\frac{\rmi}{4}}}
\newcommand{\onetwelfths}{{\textstyle\frac{1}{12}}}
\newcommand{\partialr}{\overrightarrow{\partial}}
\newcommand{\partiall}{\overleftarrow{\partial}}
\newcommand{\pfrac}[2]{\frac{\partial{#1}}{\partial{#2}}}
\newcommand{\pfracr}[2]{\frac{\partialr{#1}}{\partial{#2}}}
\newcommand{\pfracl}[2]{\frac{\partiall{#1}}{\partial{#2}}}
\renewcommand{\dfrac}[2]{\frac{\mathrm{d}{#1}}{\mathrm{d}{#2}}}
\newcommand{\ppfrac}[3]{\frac{\partial^{2}{#1}}{\partial{#2}\partial{#3}}}
\newcommand{\pppfrac}[4]{\frac{\partial^{3}{#1}}{\partial{#2}\partial{#3}\partial{#4}}}
\newcommand{\vecb}[1]{\pmb{#1}}
\newcommand{\exref}[1]{Example~\ref{#1}}
\newcommand{\pref}[1]{Part~\ref{#1}}
\newcommand{\Index}[1]{#1\index{#1}}
\newcommand{\detpartial}[2]{\left| \pfrac{#1}{#2} \right|}
\newcommand{\eref}[1]{Eq.~(\ref{#1})}
\newcommand{\fref}[1]{Fig.~\ref{#1}}
\newcommand{\sref}[1]{Section~\ref{#1}}
\newcommand{\tref}[1]{Table~\ref{#1}}
\newcommand{\dete}{\varepsilon}
\newcommand{\deteinv}{\frac{1}{\dete}}
\newcommand{\psibar}{\bar{\psi}}
\newcommand{\Psibar}{\bar{\Psi}}
\newcommand{\pibar}{\bar{\pi}}
\newcommand{\Pibar}{\bar{\Pi}}
\newcommand{\mtilde}{\mu}
\hyphenation{one-di-men-sio-nal con-fi-gu-ra-tion Ma-the-ma-ti-ca geo-metric parallel-epiped}
\normalem

\title{Dark energy and inflation invoked in CCGG by locally contorted space-time}
\author[1]{David~Vasak\thanks{vasak@fias.uni-frankfurt.de}} 
\author[1]{Johannes~Kirsch}
\author[1,2]{Jürgen~Struckmeier} 

\affil[1]{Frankfurt Institute for Advanced Studies (FIAS), Ruth-Moufang-Strasse~1, 60438 Frankfurt am Main, Germany}
\affil[2]{Goethe Universit\"at, Max-von-Laue-Strasse~1, 60438~Frankfurt am Main, Germany}
\maketitle


\begin{abstract}%
The cosmological implications of the Covariant Canonical Gauge Theory of Gravity (CCGG) are investigated. We deduce that, in a metric compatible geometry, the requirement of covariant conservation of matter invokes torsion of space-time. In the Friedman model this leads to a scalar field built from contortion and the metric with the property of dark energy, which transforms the cosmological constant to a time-dependent function. Moreover, the quadratic, scale invariant Riemann-Cartan term in the CCGG Lagrangian endows space-time with kinetic energy, and in the field equations adds a geometrical curvature correction to Einstein gravity. Applying in the Friedman model the standard $\Lambda$CDM parameter set, those equations yield a cosmological field depending just on one additional, dimensionless ``deformation'' parameter of the theory that determines the strength of the quadratic term, viz. the deviation from the Einstein-Hilbert ansatz. Moreover, the apparent curvature of the universe differs from the actual curvature parameter of the metric.

The numerical analysis in that parameter space yields three cosmology types: (I) A bounce universe starting off from a finite scale followed by a steady inflation, (II) a singular Big Bang universe undergoing a secondary inflation-deceleration phase, and (III) a solution similar to standard cosmology but with a different temporal profile. The common feature of all scenarios is the graceful exit to the current dark energy era. The value of the deformation parameter can be deduced by comparing theoretical calculations with observations, namely with the SNeIa Hubble diagram and the deceleration parameter. That comparison implies a considerable admixture of scale invariant quadratic gravity to Einstein gravity.

This theory also sheds new light on the resolution of the cosmological constant problem and of the Hubble tension.
\end{abstract}




\section{Introduction} \label{sec:Intro}
The nature of dark matter and dark energy is a long-standing unresolved problem in current cosmology. These ingredients have been added to Einstein's General Relativity (GR) in order to account for the observed accelerating expansion and the missing mass in the universe. An alternative direction to account for these observations has been to modify the Einstein-Hilbert theory by higher-order curvature terms and/or auxiliary scalar fields~\cite{wetterich88, wetterich15, starobinsky80, starobinsky82}. 
Here we consider the cosmological impact of the Covariant Canonical Gauge theory of Gravity (CCGG)~\cite{struckmeier08, struckmeier13, struckvasak15, struckmeier17a, struckmeier18a}, a classical Palatini field-theory extending Einstein gravity by a quadratic Riemann-Cartan invariant. 

\medskip
The paper is organized as follows. 
In \sref{sec:CCGG} we sketch the field-theoretical basis of the gauge theory of gravity, an approach in the spirit of earlier work on gauge theories of gravitation~\cite{utiyama56, kibble67, sciama62, hehl76, hayashi80}, but relying on the mathematical rigorousness of the canonical transformation theory in the Hamiltonian picture. That framework leads to a modified version of the Einstein equation in which torsion is admitted and metric compatibility dynamically implemented. In \sref{sec:CovCons} we discuss why torsion is a necessary degree of freedom in the theory, how the cosmological constant is promoted to a cosmological field, and how the quadratic extension in the Hamiltonian  gives rise to a \emph{geometrical stress} tensor. We do not involve any auxiliary dynamical scalar fields nor do we invoke any further ad hoc modifications of Einstein gravity. 

\medskip
The CCGG-Friedman model, in which dynamic dark energy and curvature appear as energy stores\footnote{For treating dark energy and inflation as common geometry-driven effects see also Refs.~\cite{capozziello02, capozziello03, chen09, capozziello14}} is reviewed in~\sref{sec:Friedman}, with details collected in the Appendix. 
Not only becomes the cosmological ``constant'' a time dependent quantity but also the apparent curvature while the curvature parameter of the FLRW metric remains unchanged.
The numerical solutions are discussed in~\sref{sec:scenarioanal}. By setting the priors to coincide with the~$\Lambda$CDM~parameter set, the geometry-induced corrections give rise to a time dependent dark energy function. The evolution of the universe is then determined by the new, dimensionless parameter of the theory that fixes the strength of the quadratic, scale invariant term relative to Einstein gravity (thus called the \emph{deformation parameter}).  

\medskip
The deformation parameter also influences the redshift dependence of the theoretical luminosity-distance modulus, and its comparison with the SNeIa Hubble diagram allows to gain information on its value (cf. \sref{sec:SNeIa}). Moreover, in \sref{sec:CosmConstProb} the relation of the free parameter of the CCGG theory with the value of the cosmological constant is discussed. 

\medskip
The paper closes with a summary and conclusion section.

\section{The Covariant Canonical Gauge theory of Gravity} \label{sec:CCGG}

The application of the canonical Hamiltonian transformation theory to classical relativistic matter fields has been pioneered by Struckmeier et al. and proven to derive the Yang-Mills gauge theory from first principles \cite{struckmeier08, struckmeier13}. At the heart of this framework is the requirement that the system dynamics is given by an action integral that remains invariant under prescribed local transformations of the original (matter) fields. Those transformations are implemented in the covariant de Donder-Hamiltonian formalism \cite{dedonder30} by the choice of a generating function, specifically designed for any given underlying symmetry group. 
That formalism unambiguously introduces symmetry dependent gauge fields and fixes their interaction with the matter fields. The kinetic portion of the newly introduced gauge fields, i.e. the Hamiltonian of non-interacting gravity, is not entirely determined by the gauging process, though. It is rather introduced as an \emph{educated guess} based on physical considerations and empirical insights.   

\medskip
Applying the above framework to the diffeomorphism group
paves a novel path to implementing Einstein's Principle of General Relativity to arbitrary classical relativistic systems of matter fields. 
In the resulting
first-order theory the space-time geometry is described by both, the affine connection $\gamma\indices{^{\alpha}_{\beta\mu}}$ that is not necessarily symmetric in $\beta$ and $\mu$, and the independent metric tensor $g_{\mu\nu}$\footnote{When spinors are included, the tetrad field substitutes the metric as the fundamental field of the theory, and the spin connection emerges as a gauge field.}.  The fundamental fields describing the dynamics of gravitation encompass, in addition to the connection and metric, also the ``affine momentum'' fields, the rank-4 tensor density conjugate to the affine connection,  and the ``metric momentum'' field, a rank-3 tensor density conjugate to the metric. The Hamiltonian (scalar) density, $\tilde{\HCd}_{\mathrm{Gr}}$, of space-time dynamics extends the Einstein-Hilbert ansatz by  a quadratic invariant built from the affine momentum tensor \cite{struckmeier17a, benisty18a}. In order to comply with the key observations that already gave credibility to Einstein's equation~\cite{struckmeier17a}, we set 
\begin{equation} \label{CCGE}
 \tilde{\HCd}_{\mathrm{}} = 
 \tilde{\HCd}_{\mathrm{Gr}}  + \tilde{\HCd}_{\mathrm{matter}} 
 \end{equation}
 with\footnote{Here we also consider the constant term $g_4 \sqrt{-g}$ that was omitted in \cite{struckmeier17a}.} 
 \begin{equation} \label{def:Hdyn}
 \tilde{\HCd}_{\mathrm{Gr}} =
 \frac{1}{4g_1\, \sqrt{-g}}\,\tilde{q}\indices{_\eta^{\alpha\xi\beta}}\,\tilde{q}\indices{_\alpha^{\eta\tau\lambda}}\,g_{\xi\tau}\,g_{\beta\lambda} 
 + 
 g_2\,\tilde{q}\indices{_\eta^{\alpha\eta\beta}}\,g_{\alpha\beta}
 - 
 g_4 \sqrt{-g}.
\end{equation}
The matter Hamiltonian $\tilde{\HCd}_{\mathrm{matter}}$ includes coupling of matter fields to curved space-time. 
The $(3,1)$ tensor density $\tilde{q}\indices{_\eta^{\alpha\xi\beta}} = \sqrt{-g}\,q\indices{_\eta^{\alpha\xi\beta}}$ is the affine momentum field mentioned above.  
The quadratic term endows space-time with kinetic energy and thus inertia, and fundamentally modifies its dynamics.

The coupling constants $g_1$, $g_2$ 
and $g_4$
have the dimensions $[g_1] = 1$, $[g_2] = L^{-2}$,
and $[g_4] = L^{-4}$. $g_4$ is usually identified with the vacuum energy density and gives rise to the so-called cosmological constant problem \cite{weinberg89}.
(Our conventions are the signature $(+,\,-,\,-,\,-)$ of the metric, and natural units $\hbar = c = 1$. A comma before an index denotes partial derivative, a semicolon denotes covariant derivative with the affine connection given by the gauge field. Pairs of indices in parentheses (brackets) denote (anti)symmetrization.)

\medskip
The gauging process leads to the action integral
\begin{equation} \label{actionintegral2}
S\!=
\!\int\!d^4\!x \, \tilde{\LCd}_{\mathrm{}}\!=\! 
\!\int\!d^4\!x\,\left( 
  \tilde{k}\indices{^{\alpha\nu\mu}} \, g_{\alpha\nu;\mu}  -
  \onehalf \tilde{q}\indices{_\alpha^{\beta\nu\mu}}  R\indices{^{\alpha}_{\beta\nu\mu}}
   - \tilde{\HCd}_{\mathrm{Gr}} +\tilde{\LCd}_{\mathrm{matter}} \right),
\end{equation}
where the total Lagrangian, a world scalar density, is split up into the modified gravity Lagrangian, displayed explicitly as a Legendre transform of the Hamiltonian $ \tilde{\HCd}_{\mathrm{Gr}}$ of \eref{def:Hdyn}, and the Lagrangian of matter,~$\tilde{\LCd}_{\mathrm{matter}}$, the Legendre transform of the yet unspecified~$\tilde{\HCd}_{\mathrm{matter}} $. 
In the integrand the canonical non-tensorial ``velocity'' of the affine connection, $\gamma\indices{^{\eta}_{\alpha\mu,\nu}}$, as defined in the covariant de Donder formalism, is naturally substituted by the Riemann-Cartan tensor
\begin{equation}
 R\indices{^{\eta}_{\alpha\mu\nu}} \, =
 \, \pfrac{\gamma\indices{^{\eta}_{\alpha\nu}}}{x^{\mu}} -
 \, \pfrac{\gamma\indices{^{\eta}_{\alpha\mu}}}{x^{\nu}} +
 \, \gamma\indices{^{\eta}_{\xi\mu}}\gamma\indices{^{\xi}_{\alpha\nu}} -
 \, \gamma\indices{^{\eta}_{\xi\nu}}\gamma\indices{^{\xi}_{\alpha\mu}}
\end{equation}
built from in general asymmetric affine connection coefficients. This is not an ad hoc substitution, but is the result of the requirement of diffeomorphism covariance implemented via the canonical transformation framework and the gauging process \cite{struckmeier17a}. $\tilde{k}\indices{_{\alpha\beta\nu}}$ is the metric momentum, the field conjugate to the metric~$g_{\alpha\beta}$. It is by definition symmetric in its first two indices.

\medskip
Variation of \eref{actionintegral2}  with respect to the affine momentum gives the canonical equation 
\begin{equation} \label{eq:r}
 R\indices{^{\alpha}_{\beta\nu\mu}} = -2 \pfrac{\tilde{\HCd}_{\mathrm{Gr}}}{\tilde{q}\indices{_\alpha^{\beta\nu\mu}}}.
\end{equation}
Its solution is for the Hamiltonian of~\eref{def:Hdyn}
\begin{equation} \label{def:momentum}
 q_{\eta\alpha\xi\beta} = g_1\,\left(R_{\eta\alpha\xi\beta} - \hat{R}_{\eta\alpha\xi\beta}\right),
\end{equation}
where 
\begin{equation} \label{def:maxsymR}
\hat{R}_{\eta\alpha\xi\beta} = g_2 \left(g_{\eta\xi}\,g_{\alpha\beta} - g_{\eta\beta}\,g_{\alpha\xi} \right)
\end{equation}
is the Riemann curvature tensor of the maximally symmetric space-time, the (Anti) de Sitter geometry with the Ricci scalar curvature $12\,g_2$. The affine momentum of space-time thus accounts for deformations of the space-time geometry relative to the (A)dS ground state.

\medskip
Notice that the canonical field equation for the metric momentum,
\begin{equation} \label{eq:Dg/dx}
g_{\alpha\beta;\nu} =  \pfrac{\tilde{\HCd}_{\mathrm{Gr}}}{\tilde{k}\indices{^{\alpha\beta\nu}}} = 0,
\end{equation}
implements dynamically metric compatibility. This is due to the fact that the kinetic Hamiltonian $\tilde{\HCd}_{\mathrm{Gr}}$,  \eref{def:Hdyn}, has been chosen to be independent of $\tilde{k}\indices{^{\alpha\beta\nu}}$, the ``metric momentum'' conjugate to the metric tensor\footnote{By Legendre transformation $\tilde{k}\indices{^{\alpha\beta\nu}}$ becomes then a Lagrange multiplier in the Lagrangian density.}.

\medskip
The so-called consistency (``CCGG'') equation that extends Einstein gravity~\cite{struckmeier17a, struckmeier18a} is obtained as a combination 
of the canonical equations. It can be written as a \emph{local} balance equation,  
\begin{equation} \label{eq:modEinstein}
 -\Theta\indices{^\mu^\nu} = T\indices{^\mu^\nu} \\
\end{equation}
with
\begin{subequations}
 \begin{alignat}{4}
  \Theta\indices{^\mu^\nu} &:= \frac{2}{\sqrt{-g}}\,\pfrac{\tilde{\HCd}_{\mathrm{Gr }}}{g\indices{_{\alpha\beta}}} \label{def:Q0} \\
       T\indices{^\mu^\nu} &:= \frac{2}{\sqrt{-g}}\,\pfrac{\tilde{\HCd}_{\mathrm{matter}}}{g\indices{_{\alpha\beta}}},
 \end{alignat}
\end{subequations}
and is similar to the stress-strain relation in elastic media. In analogy to the the energy-momentum (``stress-energy'') tensor of matter, $T\indices{^\mu^\nu}$, we interpret $\Theta\indices{^\mu^\nu}$ as the energy-momentum (``strain-energy'') tensor of space-time\footnote{This also implies that the total energy of the universe is zero, consistent with Jordan's conjecture, cf. \cite{jordan39}. Taking the vacuum expectation value of the "quantum analogue" of this equation, we would expect the vacuum energy densities of gravity and the matter to cancel each other, perhaps up to some residual value that can be identified with $g_4 \approx 0$.}.  Calculating now the strain-energy tensor \eqref{def:Q0} with the CCGG Hamiltonian~\eqref{def:Hdyn}, and substituting \eref{eq:r} for the momentum tensor, gives 
\begin{equation} \label{def:Theta2}
 \Theta\indices{^\mu^\nu} = 
 -g_1 \, Q\indices{^\mu^\nu}
           - 
           2g_1g_2\, \left(G\indices{^\mu^\nu} 
           - 
           3g_2\,g\indices{^\mu^\nu} \right) 
           - 
           g_4\,g\indices{^\mu^\nu}.
\end{equation}
where  
\begin{equation}
G\indices{^\mu^\nu} := R\indices{^{(\mu\nu)}}
           - \onehalf g\indices{^\mu^\nu} \, R
\end{equation}
is the Einstein tensor\footnote{The Einstein tensor as derived from the canonical equations of motion contains only the symmetrized Ricci tensor. While in the absence of torsion the Ricci tensor is symmetric, it is not the case for non-zero torsion. The anti-symmetric Ricci tensor interacts with the spin density of matters.}, and  
\begin{equation}
Q\indices{^\mu^\nu} :=  R^{\alpha\beta\gamma\mu}\, R\indices{_{\alpha\beta\gamma}^{\nu}}
           - \quarter g\indices{^\mu^\nu} \, R^{\alpha\beta\gamma\xi}\, R_{\alpha\beta\gamma\xi},
\end{equation}
is a trace-free, (symmetric) quadratic Riemann-Cartan concomitant.  \eref{def:Theta2} is a generalization of the l.h.s. of the Einstein equation in three aspects. Firstly, the Palatini formalism is used, so the affine connection and the metric are still independent fields, torsion of space-time is admitted, and a quadratic Riemann-Cartan term added. In the Lagrangian constructed by Legendre transformation that term is built from the Kretschmann scalar $R^{\alpha\beta\mu\nu}\, R_{\alpha\beta\mu\nu}$.

\section{Geometrical stress energy and Cartan contortion density} \label{sec:CovCons}
In the CCGG formalism, the tensors, $Q\indices{^\mu^\nu} $ and $G\indices{^\mu^\nu} $, are not necessarily covariantly conserved. That can even be the case if we set torsion to zero, as is known for Palatini type theories with torsion not a priori excluded \cite{fabbri14,benisty18a}. Indeed, straight algebra\footnote{We thank Julia Lienert for checking this identity with Maple.} shows that
\begin{equation} \label{covderQ}
Q\indices{_{\nu}^{\mu}_{;\mu}} \equiv  R\indices{^{\alpha\beta\gamma\mu}_{;\mu}} \, R\indices{_{\alpha\beta\gamma}_\nu}
+ R\indices{^{\alpha\beta\gamma\mu}} \left(
R\indices{_{\alpha\beta\nu\tau}} \, S\indices{^{\tau}_{\gamma\mu}} - 
2 R\indices{_{\alpha\beta\gamma\tau}} \, S\indices{^{\tau}_{\nu\mu}} \right),
\end{equation}
where
\begin{equation} \label{def:torsiontensor}
 S\indices{^\lambda_{\mu\nu}} = \onehalf(\gamma\indices{^{\lambda}_{\mu\nu}}- \gamma\indices{^{\lambda}_{\nu\mu}})
\end{equation}
is Cartan's torsion tensor, vanishes for zero torsion and metric compatibility only if
\begin{equation} \label{covderR}
\bar{R}\indices{_{\alpha\beta\gamma}^\nu} \, \bar{\nabla}_\mu \bar{R}\indices{^{\alpha\beta\gamma\mu}}  = 0.
\end{equation}
The overbared quantities are calculated using the Christoffel symbol,  
\begin{equation}\label{eq:gammaLC}
\gamma\indices{^{\lambda}_{\mu\nu}} \rightarrow 
\genfrac{\lbrace}{\rbrace}{0pt}{0}{\lambda}{\mu\nu},
\end{equation}
as for metric compatible and torsion-free geometries the affine connection must be the Christoffel symbol uniquely determined by the metric via the Levi-Civita relation. 

\medskip
It is important to stress that the covariant conservation law  for the strain-energy tensor, \eref{covderQ}, is \emph{not} an identity, yet it facilitates, in addition to metric compatibility, a constraint linking the metric and the affine connection.  The requirement \eqref{covderR} then restricts or even fixes the metric, and, in addition, implies that the r.h.s. of the consistency equation \eqref{eq:modEinstein}, the stress-energy tensor, is covariantly conserved, too.

\medskip
Vice versa, requesting the stress-energy tensor of matter to be covariantly conserved, might in the CCGG theory lead to the necessity to adjust the affine connection with the given metric beyond the Levi-Civita relation. Physically this opens a new channel within the dynamical geometry to take up its deformation energy, and this channel is based on torsion. 

\medskip
This can be illustrated as follows. If in the classical, macroscopic limit, torsion is neglected and the affine connection is assumed to be the Christoffel symbol, then \eref{eq:modEinstein} becomes
\begin{equation} \label{eq:modEinstein2}
 g_1\left( \bar{R}^{\alpha\beta\gamma\mu}\, \bar{R}\indices{_{\alpha\beta\gamma}^\nu}
           - \quarter g^{\mu\nu} \, \bar{R}^{\alpha\beta\gamma\delta}\, \bar{R}_{\alpha\beta\gamma\delta}\right)
           + \frac{1}{8\pi G}\, \left[ \bar{R}\indices{^\mu^\nu}
           - g^{\mu\nu} \left( \onehalf \bar{R} + \lambda_0 \right)\right] = \bar{T}\indices{^\mu^\nu}.
\end{equation}
Notice that for aligning with the syntax used in GR, the coupling constants $g_i$ in \eref{eq:modEinstein2} have been expressed in terms of the gravitational coupling constant $G$ and the cosmological constant $\lambda_0$ as follows:
\begin{subequations}
\begin{alignat}{4}
   g_1\,g_2 &\equiv 
   \frac{1}{16\pi G} = 
   \onehalf M_p^2 \label{def:constantsg1}\\
   6g_1\,g_2^2 
   - 
   g_4 &\equiv \frac{\lambda_0}{8\pi G} \; = M_p^2\,\lambda_0.
   \label{def:constantsg3}
\end{alignat}
\end{subequations}
$M_p := \sqrt{1/8\pi G}$ is the reduced Planck mass. 
Combining the above equations yields
\begin{equation} \label{def:constantsg32}
 \lambda_0 = 
 3g_2 
 - 
 8\pi G\,g_4 = \frac{1}{M_p^2} \left( \frac{3 M_p^4}{2 g_1}
 - 
 g_4 \right)
 ,
\end{equation}
i.e.\ we find the cosmological constant being generated by the (A)dS curvature of the ground state of space-time, and the vacuum energy~$g_4$ \cite{vasak19a}. The parameter~$g_1$ is the \emph{deformation parameter} of the theory\footnote{This result reminds of earlier approaches under the heading of de Sitter relativity to derive the cosmological constant and to explain cosmic coincidence and time delays of extra-galactic gamma-ray flares (see for example~\cite{aldrovandi09}).} as it determines the strength of the quadratic Riemann-Cartan extension of Einstein gravity.  
(The coupling constant~$g_2 = M_p^2/2g_1$ is thus the inverse of the deformation parameter.)

\medskip
In order to explicitly work out the differences between the standard, GR based cosmology and the CCGG model, we request the stress-energy tensor to be covariantly conserved: 
\begin{equation} \label{eq:covderTbar}
 \bar{\nabla}_\nu \bar{T}\indices{^\mu^\nu} = 0.
\end{equation}
This is consistent as long as 
\begin{equation}
 \bar{\nabla}_\nu \bar{\Theta}\indices{^\mu^\nu} = 0
\end{equation}
holds, which for torsion-free geometries reduces to \eref{covderR} for the Riemann tensor. 

\medskip
If a specific ansatz for the metric under the additional assumption of zero-torsion solves the over-bared~\eref{eq:modEinstein2}, but fails to satisfy the constraint~\eqref{covderR}, the affine connection must not be Christoffel, though. With metricity in place, the most general form of the affine connection is
\begin{equation}\label{eq:gammaing}
\gamma\indices{^{\lambda}_{\mu\nu}} =
\genfrac{\lbrace}{\rbrace}{0pt}{0}{\lambda}{\mu\nu}+
K\indices{^{\lambda}_{\mu\nu}} 
\end{equation}
with the contortion tensor\footnote{The generalization is to take also non-metricity into account. The autor of \cite{queiruga19} considers a symmetric affine connection with non-metricity.}
\begin{equation} \label{def:contortion}
K\indices{_{\lambda}_{\mu\nu}} \,=\,
S\indices{_{\lambda\mu\nu}} - S\indices{_{\mu\lambda\nu}} + S\indices{_{\nu\mu\lambda}}\,=\,-K\indices{_{\mu}_{\lambda\nu}}
. 
\end{equation}
The contortion tensor is a linear combination of the torsion tensor \eqref{def:torsiontensor}, and the metric. Obviously, deviating for a given metric from the Levi-Civita ansatz for the connection is equivalent to introducing torsion of space-time. 

\medskip
The Riemann-Cartan tensor can now be split into metric (Riemann) and torsion-dependent (Cartan) portions,
\begin{equation}
 R\indices{_{\alpha\beta\gamma\sigma}}(\gamma\indices{^{\lambda}_{\mu\nu}}) 
 \equiv \bar{R}\indices{_{\alpha\beta\gamma\sigma}} + P\indices{_{\alpha\beta\gamma\sigma}},
\end{equation}
where 
\begin{equation} \label{def:Ptensor}
P\indices{_{\lambda}_{\sigma\mu\nu}} 
 := \bar{\nabla}_\mu K\indices{_{\lambda}_{\sigma\nu}} -
    \bar{\nabla}_\nu K\indices{_{\lambda}_{\sigma\mu}} -
    K\indices{_{\lambda}_{\beta\nu}} K\indices{^{\beta}_{\sigma\mu}} + K\indices{_{\lambda}_{\beta\mu}} K\indices{^{\beta}_{\sigma\nu}} 
\end{equation}
denotes the \emph{Cartan curvature tensor} which is antisymmetric in the first and the second pair of indices. 

\medskip
Then with the definition of the symmetric $(2,0)$ tensor,
 \begin{align}
 &\xi^{\mu\nu}(g,S) := \, \bar{Q}^{\mu\nu} - Q^{\mu\nu} \\
 &\quad = \left(\quarter g^{\mu\nu}\,\delta^\tau_\sigma - g^{\nu\tau}\,\delta^\mu_\sigma \right)
 \left( \bar{R}\indices{^{\alpha\beta\gamma\sigma}} P\indices{_{\alpha\beta\gamma\tau}}
 + \bar{R}\indices{_{\alpha\beta\gamma\tau}} P\indices{^{\alpha\beta\gamma\sigma}}
 + P\indices{^{\alpha\beta\gamma\sigma}} P\indices{_{\alpha\beta\gamma\tau}} \right), \nonumber
 \end{align}
 and the decomposition
\begin{equation}
G\indices{^\mu^\nu} = \bar{G}\indices{^\mu^\nu} + P^{(\mu\nu)}
           - \onehalf \,g^{\mu\nu}\,  P, 
\end{equation}
the overbared CCGG equation \eqref{eq:modEinstein2} is corrected to
\begin{equation} \label{eq:modEinstein4}
 g_1\, \bar{Q}^{\mu\nu} 
           + \frac{1}{8\pi G}\,\left[ 
           \bar{G}^{\mu\nu} - g^{\mu\nu} \Lambda(x) \,\right]  
           = \bar{T}^{\mu\nu} - \frac{1}{8\pi G} P'^{(\mu\nu)}
           + g_1 \xi^{\mu\nu}.
\end{equation}
The deviation from \eref{eq:modEinstein2} is composed of terms that vanish with vanishing torsion\footnote{Notice that for an application in cosmology with just classical matter the corresponding stress tensor is independent of the affine connection and hence is independent of torsion, $\bar{T}^{\mu\nu} \equiv T^{\mu\nu}$.}. Thereby 
\begin{equation}
P'^{(\mu\nu)} = P^{(\mu\nu)} - \quarter g^{\mu\nu} P 
\end{equation}
is the trace-free Cartan-Ricci tensor. All tensors in this equation, including~$\xi^{\mu\nu}$, are symmetric by definition. 

\medskip
When placing the tensors $P'^{(\mu\nu)}$ and $\xi^{\mu\nu}$ next to the stress-energy tensor of matter on the r.h.s. of the equation, it appears as a new, \emph{geometrical stress tensor}. The geometrical stress-energy tensor is not covariantly conserved, and energy transfer from space-time is possible.
Moreover, by its very definition, the term $P'^{(\mu\nu)}$ on the r.h.s. of \eref{eq:modEinstein4} is trace-free, like the energy-momentum tensor of radiation or relativistic matter. Leaving only the Einstein tensor and the cosmological term on the l.h.s. of the equation, i.e. taking the \emph{Einstein view} of the system where all geometrical terms are considered as part of the extended, then covariantly conserved, stress-energy tensor, enables to study the newly emerging phenomena in relation to General Relativity. 

\medskip
The Cartan-Ricci curvature scalar, $P(x)$, that we call \emph{Cartan contortion density} as it is built from contortion and metric, is combined with the cosmological constant to the scalar \emph{cosmological field}
\begin{equation} \label{def:cosmfield}
  \Lambda(x) := \lambda_0 + \quarter P(x). 
\end{equation}
$\Lambda(x)$ reduces to a constant in torsion-free geometries, and may not vanish even if the cosmological constant $\lambda_0$ does. 
In the following we demonstrate that, under simplifying assumptions in the Friedman model, a unique solution of the cosmological field exists.


\section{The CCGG-Friedman model} \label{sec:Friedman}

The Friedman model universe~\cite{friedman22,weinberg72} is endowed with the Friedman-Lema\^{i}tre-Robertson-Walker (FLRW) metric with curvature characterized by the parameter $K_0$,
\begin{equation}\label{def:RWmetric}
ds^2 = dt^2 - a^2(t) \left[\frac{dr^2}{1-K_0 r^2} + r^2\left(d\theta^2 + \sin^2(\theta)\,d\varphi^2\right)\right].
\end{equation}
The dimensionless parameter $a(t)$, the relative scale of the spatial section of the metric as function of the cosmological time $t$, remains the only dynamical freedom left. If $t_0$ is the current age of the universe, $a(t_0) = 1$ applies to today. The parameter $K_0$ fixes the type of the underlying geometry: $K_0 = 0$ flat, $K_0 > 0$ spherical, $K_0 < 0$ hyperbolic. 

\medskip
With this one-parameter FLRW metric ansatz, \eref{covderR} is satisfied only for three non-constant solutions for the scale function $a(t)$ 
which, for cosmology, is in general too restrictive as it is independent of the matter content of the universe! Hence~\eref{eq:modEinstein4} must be considered with the tensor corrections as outlined above.  
The torsion-induced tensor corrections on the r.h.s., $P'^{(\mu\nu)}$ and $\xi^{\mu\nu}$, must be diagonal, mimicking radiation and matter, respectively, with some unknown equations of state\footnote{Whether these terms can explain the effect of (hot and cold) dark matter, and how they might impact cosmological perturbations, is a topic for separate studies.}. 
We accommodate schematically the tensor corrections to the stress-energy tensor by including cold dark matter in the density of dust, and neglect radiation-like contributions (aka hot dark matter). Only the scalar Cartan contortion density in the cosmological field  will be retained on the strain-energy side of the equation as a yet unknown dynamical quantity. 

\medskip
Due to the isotropy and homogenity of the FLRW geometry that cosmological field~$\Lambda(x)$ can only depend on the universal time~$t$. 
It formally corresponds to a density with the equation of state of dark energy\footnote{
The impact of the torsion-related corrections of the Einstein equation on cosmology has been discussed in \cite{minkowski86}. Cosmology with a homogeneous spin density (aka Weyssenhoff fluid) were addressed in Refs. \cite{obukhov87, obukhov93, boehmer06, brechet07, poplawski18, unger19}. 
A time dependent cosmological constant has been also derived from string theory~\cite{basilakos19}, and by the renormalization group method~\cite{myrzakulov15}.}, 
and is therefore called here the \emph{dark energy function}. The analysis is further simplified by adopting the scaling ansatz\footnote{
Under the assumption that $a(t)$ is a strictly monotonical function, $t(a)$ exists and is well defined. In case of a bouncing or oscillating universe, though, care must be taken and the branches with $\dot{a} > 0$ and $\dot{a} < 0$ considered separately.}
\begin{equation} \label{def:f(a)}
\Lambda(t) = \Lambda(t(a)) =: \Lambda_0 \, f(a)
\end{equation}
with the dimensionless function $f(a)$, and  
the yet unspecified constant~$\Lambda_0$.

\medskip
The definitions and algebra leading to the Friedman equations are 
    given in Appendix~\ref{App:FM}. The result is the modified Hubble function,
 \begin{align} \label{eq:modFriedman4}
 H^2(a) &= 
 \sum_{i=r,m}\,C_i\,a^{-n_i} - \frac{k(a)}{a^2}+ C_\Lambda\,f(a),
\end{align}
where the geometrical effects emerging from the quadratic term have been combined with the curvature parameter $K_0$ of the FLRW metric to the \emph{curvature function}\footnote{Treating the term invoked by the quadratic extension of the Hamiltonian as a curvature correction might seem arbitrary. However, its origin is the space-time side of the equation, hence the other sensible option would be to combine it with the dark energy function. For an early analysis of that combination with a slightly different interpretation of the correction terms see~\cite{vasak19a}.}
\begin{equation} \label{def:chi}
 k(a) := K_0 -
 \frac{
 \left( \quarter C_m  a^{-3} + C_\Lambda \,f(a) \right) \left( \threequarter C_m \, a^{-1} + C_r\,a^{-2} \right)
 }
 {
 \onehalf g_2 - \quarter C_m\,a^{-3} - C_\Lambda \,f(a)
 }.
 \end{equation}
$k(a)$ is well defined since the function $f(a)$ obeys the unique differential equation derived in Appendix \ref{App:f(a)}:
\begin{align} \label{ODE:f(a)}
 \frac{df}{da} &=
  \frac{3C_m}{4C_\Lambda} a^{-4} \, \times \\
  &\frac{
  \onehalf g_2 \left(\threequarter C_m a^{-3} + C_r a^{-4}\right)
  - \left(\onehalf g_2 - \quarter C_m a^{-3} - C_\Lambda \,f \right)
  \left(\quarter C_m a^{-3} + C_\Lambda \,f \right)
    }
  {\onehalf g_2 \left(\threequarter C_m a^{-3} + C_r a^{-4}\right)
    +   \left(\onehalf g_2 - \quarter C_m a^{-3} - C_\Lambda \,f \right)^2
  }. \nonumber
\end{align}
By setting $g_1 = 0$ (which means $g_2 = \infty$) and $f(a) \equiv 1$ we recover the Einstein-Friedman equation for the Hubble function based on General Relativity.  

Moreover, as shown in the Appendix~\ref{App:f(a)}, the modified Friedman equations cannot be solved with $f(a) = const.$, confirming the necessity to include the torsion corrections enforced by requiring concomitantly the FLRW metric and the covariant conservation of the energy-momentum tensor of matter. 

\medskip
If we require the dark energy term to reproduce the observed present-day value of the cosmological constant,  and set $\Lambda_0 = \Lambda_{\mathrm{obs}}$, the initial condition~$f(1)=1$ must hold. 
Similar reasoning for the curvature term gives $k(1) = k_{\mathrm{obs}}$, which implies
 \begin{equation} \label{eq:k0}
 k_{\mathrm{obs}} \equiv k(1) = K_0 - 
 \frac{
 \left( \quarter C_m  + C_\Lambda \right) \left( \threequarter C_m + C_r \right)
 }
 {
 \onehalf g_2 - \quarter C_m - C_\Lambda
 }.
 \end{equation}
 Setting $\Delta k := K_0 - k_{\mathrm{obs}}$, this relation can be resolved for $g_2$ giving
 \begin{equation} \label{prooffne1}
  \onehalf \, g_2(\Delta k) = \frac{1}{\Delta k} \, \left(
  \quarter\,C_m+C_\Lambda\right) \left(\threequarter \,C_m +C_r+\Delta k  \right).
 \end{equation}
With \eref{def:constantsg1}, an equivalent expression for $g_1$ is obtained.
Obviously, $g_2$ diverges and $g_1$ vanishes for $K_0 \rightarrow k_{\mathrm{obs}}$. For a dark energy-dominated system, i.e. with $C_m = C_r = 0$, \eref{ODE:f(a)} yields the solution $f(a) \equiv 1$. Then \eref{eq:k0} gives  $\Delta k := K_0 - k_{\mathrm{obs}} = 0$, and we obtain $g_1 = 0$. Hence a dark energy-dominated CCGG universe is identical to the 
Einstein-Friedman model. \eref{eq:modFriedman4} is thus the equation of motion of the CCGG cosmology in the Einstein view.

The construction of the functions~$f(a)$ and~$k(a)$, and the selected boundary conditions for $a=1$, give with the \eref{eq:modFriedman4} the Hubble constant
\begin{equation}
  H_0^2 = \sum_{i=r,m,\Lambda}\,C_i - k_{\mathrm{obs}},
  \label{eq:modFriedman43}
\end{equation}
which coincides with the standard value. Hence we can adopt the parameters of the conventional Einstein-Friedman model based on the assumptions of almost flat FLRW metric, standard model of elementary particles, cold dark matter, and a constant cosmological constant~$\Lambda_0 = \Lambda_{\mathrm{obs}}$. (Concordance Model with the $\Lambda$CDM parameter set as listed in~\eref{def:constants} and Table~2.)

\begin{figure}[H]
\includegraphics[width=\linewidth]{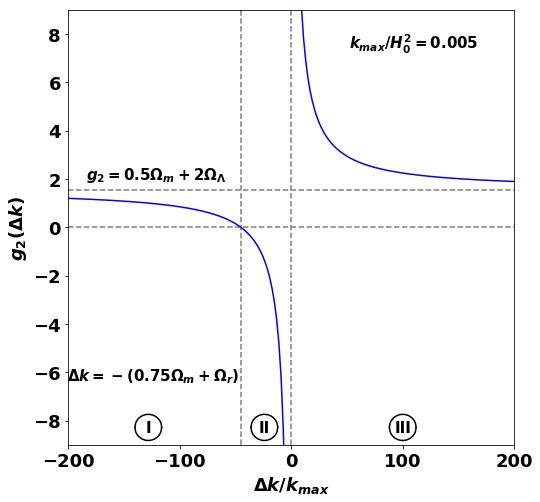}
 \caption{\footnotesize 
 The relation $g_2(\Delta k)$ is displayed in units $H_0^2$. As $g_2$ must be non-zero and finite in order to maintain the consistency of the theory, the root $\Delta \bar{k} \equiv - 3 C_m / 4 - C_r$, where $g_2 = 0$, and $\Delta k = 0$ with $g_2 \rightarrow \pm\infty$, are both ``forbidden'' values. 
 For $\Delta k  \rightarrow \pm\infty$ we find $g_2  \rightarrow C_m/2 + 2C_\Lambda$. 
 }
\label{fig:g_2(K_0)}
\end{figure}

\medskip
An important astronomical observable is also the dimensionless deceleration function
\begin{equation} \label{eq:deceleration}
q := -\frac{\ddot{a}}{\dot{a}^2}\,a \equiv -\frac{\ddot{a}}{a}\,\frac{1}{H^2} = \frac{2a^2\,M+V}{V+K_0},
\end{equation}
which explicitly depends on the curvature parameter $K_0$, and implicitly on the dark energy and curvature functions in the scale potential $V(a)$,~\eref{V(a)}. $M(a)$, viz.~\eref{def:M1}, is the Ricci scalar of the FLRW 
metric\footnote{
\begin{equation*} 
 R^{FLRW} = 6\left[\frac{\ddot{a}}{a} + \left(\frac{\dot{a}}{a}\right)^2 + \frac{K_0}{a^2}\right] = 12\,M(a) = 3 \, C_m a^{-3} + 4\, \Lambda(a) 
\end{equation*}}. 
For the $\Lambda$CDM parameter set (cf.~) the present-day deceleration parameter~$q_0 \equiv q(1)$ is
\begin{equation}\label{eq:decelerationf0}
q_0 
 \approx -0.5 + K_0/H_0^2.
\end{equation}
Obviously, the value of both, $K_0/H_0^2$ and $g_1$, is restricted by the 
measurement accuracy of $q_0$\footnote{For a discussion see for 
example~\cite{planck15, bernal17}.}. Recall that the parameter $K_0$ determines 
via \eref{prooffne1} the coupling constant $g_2$, visualized 
in~\fref{fig:g_2(K_0)}, and via $g_1 = -M_p^2/2g_2$ also the deformation 
parameter of the theory.  

\medskip
It is important to stress at this point  that $K_0 = k_{\mathrm{obs}}$ is 
possible if and only if $g_1 = 0$, i.e.\ only in the realm of the 
Einstein-Friedman model. The limiting process,~$g_1 \rightarrow 0$, is 
continuous but not convergent, though, since $g_2 \sim -1/g_1$ diverges 
there!\footnote{That would give rise to an inconsistency as the quadratic term 
then diverges in the Hamiltonian while it vanishes in the Lagrangian. $g_1$ and 
$g_2$ must thus be finite in order for the Hamiltonian theory to be consistent 
with the action principle. The  dimensionless deformation parameter $g_1$ 
appears as the measure of the \emph{inertia of space-time} in line with the 
formal r\^{o}le of the Riemann-Cartan tensor as the covariant "velocity" of the 
affine connection field.}.

\medskip
Obviously, for this type of analysis and within the present accuracy of observations the CCGG-Friedman model with the six priors of the Concordance Model is a reasonable ansatz. 

\section{Scenario analysis} \label{sec:scenarioanal}

In this Section we investigate the impact of the remaining free parameter, the curvature $K_0$ of the Friedman metric, on the evolution of the universe. 
We follow the standard practice and set $C_k = -k_{\mathrm{obs}} = 0$, i.e. we accept the present universe to be with a high accuracy flat for consistency with the observed CMB radiation isotropy~\cite{planck15}.
Then $\Delta k \equiv K_0$. We vary the parameter~$K_0 / k_{\mathrm{max}} \in \RB$ that enters the formula for the calculation of the coupling constant $g_2$ --- and the deformation parameter $g_1$. 
In order to remain close to the Concordance Model we vary $K_0$ on the scale of $k_{\mathrm{max}}$ inferred from observations, namely $|k_{\mathrm{obs}}| \le k_{\mathrm{max}} = 0.005 H_0^2$, cf.~\eref{def:kexp}. The values of $K_0$, $g_2$ (and $g_1$) applied in the following calculations are listed in~Table~1.

\medskip
For a given $K_0$ we calculate $g_2$ from~\eref{prooffne1}, and then numerically solve the differential equation~\eqref{ODE:f(a)} for~$f(a;g_2(K_0))$ using the six priors of the Concordance Model.  The dynamics of the universe's expansion is thereby determined by the equation 
\begin{equation} \label{DG1}
 \dot{a}^2 + V(a) = -K_0,
\end{equation}
derived in the Appendix~\ref{App:f(a)}, cf.~\eref{DG}. It is formally the total energy of a fictitious classical, non-relativistic point particle with mass~$2$, moving with the velocity~$\dot{a}$, in the one-dimensional potential~$V(a)$. Having the total energy $-K_0$, that particle will accelerate when "sliding down" the potential wall until it hits the minimum, and decelerate when climbing up its walls as long as its kinetic energy remains positive.  The equation $V(a_i) = -K_0$ thus determines a turning point that corresponds to a bounce of the universe at a (possibly finite) extension scale $a_i$. The area~$\{a:V(a;g_2(K_0) < -K_0\}$ is a  ``forbidden zone''\footnote{All energy condition would be violated.}. Notice that the equation of motion \eqref{DG1} is time-reversal invariant, so if $\dot{a}$ is a solution then also $-\dot{a}$, and expansion and contraction are in principle equally possible. The two branches can join continuously at the point $a$ where $\dot{a} = 0$.

\begin{table}[H]
\label{tab:Parameter1}
 \begin{tabular}[t]{|c|c|c|} 
\toprule
$K_0/k_{\mathrm{max}}$ & $g_2/H_0^2$ & $g_1$ \\
\midrule
Region I & & \\
\midrule
$-80$ & $0.66$ & $ 1.99 \times 10^{120}$ \\
$-70$ & $0.54$ & $ 2.45 \times 10^{120}$ \\
$-60$ & $0.37$ & $ 3.56 \times 10^{120}$ \\
$-46$ & $0.03$ & $ 3.85 \times 10^{121}$ \\
\midrule 
Excluded & & \\
\midrule
$-45$ & $0.00$ & $\pm \infty$ \\
\midrule 
Region II & & \\
\midrule
$-44$ & $  -0.04$ & $-3.68 \times 10^{121}$ \\
$-30$ & $  -0.78$ & $-1.67 \times 10^{120}$ \\
$-10$ & $  -5.43$ & $-2.39 \times 10^{119}$ \\
$ -1$ & $ -68.21$ & $-1.90 \times 10^{118}$ \\
\midrule 
Excluded & & \\
\midrule
$  0.00$ & $\pm \infty$ & $0.00$ \\
\midrule
Region III & & \\
\midrule
$ 1$ & $71.31$ & $1.82 \times 10^{118}$ \\
$10$ & $ 8.53$ & $1.52 \times 10^{119}$ \\
$45$ & $ 3.10$ & $4.18 \times 10^{119}$ \\
$80$ & $ 2.42$ & $5.35 \times 10^{119}$ \\
\bottomrule
\end{tabular} 
\caption{\footnotesize 
 The table lists the values of the parameter $K_0/k_{\mathrm{max}}$ used alongside the Default $\Lambda$CDM parameter set in the following calculations. Three combinations with the sign of the pertinent coupling constant $g_2(K_0)$ (in units $H_0^2$) and the deformation parameter $g_1$ (dimensionless) yield three  parameter Regions I (top), II (middle), and III (bottom). The values $g_2 = 0, \pm \infty$ are limiting cases outside the realm of the CCGG theory. $k_{\mathrm{max}} \equiv +0.005 \, H_0^2$. 
 }
 \end{table}

\medskip
In order to check the validity of the energy conditions \cite{capozziello14} in the Einstein view, the effective total energy density and pressure must include matter, radiation, dark energy and other geometry driven terms. They, and the resulting EOS, are derived from Eqs. \eqref{eq:F2} and \eqref{eq:F3}, using the definitions $\Omega_i \equiv C_i/H_0^2$ and $\rho_{crit} \equiv 3(H_0\,M_p)^2$:
\begin{subequations} \label{def:corrections}
\begin{alignat}{4}
\rho(a) :=& \,\rho_{crit}\,\left( \Omega_m\,a^{-3}+\Omega_r\,a^{-4}+\Omega_\Lambda\,f(a) - \frac{k(a)}{H_0^2\,a^{2}} \right) = \rho_{crit}\,H^2(a)/H_0^2, \\
   p(a) :=& \,\rho_{crit}\,\left( \onethird \, \Omega_r \,a^{-4} -\Omega_\Lambda\,f(a) - \onethird \,\frac{k(a)-2K_0}{H_0^2\,a^{2}}  \right), \\
\omega(a) :=& \,\frac{p(a)}{\rho(a)} = \onethird \left[ 2q(a) - 1\right].
\end{alignat}
\end{subequations}
Both, the total density and pressure reduce to the standard Einstein-Friedman expressions with $f(a) \equiv 1$ and $g_1 = 0$. Notice that the conditions NEC, WEC and DEC are satisfied if $\rho(a) \ge 0$ and $\omega(a) \ge -1$ hold. For SEC to be met $\omega(a) \ge -\onethird$ is required.

\medskip
The role of the parameter $K_0$ is intricate as its value determines implicitly --- via the coupling constant $g_2$ --- the shape of the scale potential.    
Its physical impact is analyzed in the following sections for the Regions~I~($K_0 < \bar{K}_0$), 
II~($\bar{K}_0 < K_0 < 0$), and~III~($K_0 > \bar{K}_0$), separated by the critical values~$\bar{K}_0 = -\threequarter C_m + C_r \approx -45.011 \, k_{\mathrm{max}} = 0.22506\,H_0^2$ for the Default parameter set (cf. Table~2), and $K_0 = 0$. Recall that $\bar{K}_0$ is critical as for this value $g_2$ vanishes and the deformation parameter $g_1$ is infinite which means that the CCGG equation is dominated by the scale invariant quadratic gravity. For $K_0 = 0$, on the other hand, the deformation parameter $g_1$ vanishes, corresponding to the pure Einstein gravity limit. The deformation parameter thus interpolates the theory between the scale invariant and the Einstein-Hilbert versions. The Lagrangians for both limits exist but outside of the domain of the CCGG theory.

\medskip
A remark is due on the relation of the curvatures $K_0$ vs. $k(a)$. The former is the constant parameter of the FLRW metric, the latter is the CCGG curvature function. Once $K_0$ is inferred from observations (e.g. of $q_0$) then the function $k(a;K_0)$ is fixed. On the other hand, analyzes of observations based on Einstein gravity will, at any given past or future instant $t_f$, misinterpret the value of $k(a(t_f))$ as a (possibly time-dependent) FLRW curvature parameter, albeit the space-time geometry is defined with $K_0$.

\medskip
We note in passing that we carry out the calculation with a high precision toolkit\footnote{\textcolor{black}{All calculations were performed using \emph{mpmath}, a Python library for floating point arithmetic with arbitrary precision. \cite{johansson18}}} to provide an initial evaluation and interpretation of the emerging dynamical scenarios, even though for very small scale parameters $a \rightarrow 0$ both the physics and the numerics of the results may be questioned. We leave the ultimate judgement on the physical significance of the results to a more comprehensive analysis.

\subsection{Region I: Non-singular Bounce with steady inflation} \label{sec:RegionI}

\begin{figure}[H]
\begin{minipage}{14pc}
 \includegraphics[width=14pc]{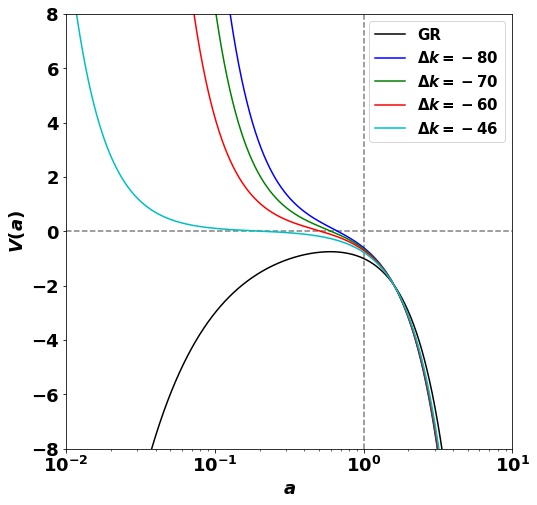}
 \caption{\footnotesize 
 The log-lin plot of the scale potentials $V(a;g_2(K_0))$ in units of $H_0^2$. The negative branch is the forbidden region of scale expansion.
 The curve labeled GR shows the potential of the conventional, flat Einstein-Friedman cosmology where $f(a) \equiv 1, g_1 = 0$. }
 \label{fig:IV(a)}
 \end{minipage}\hspace{1pc}%
 \begin{minipage}{14pc}
 \includegraphics[width=14pc]{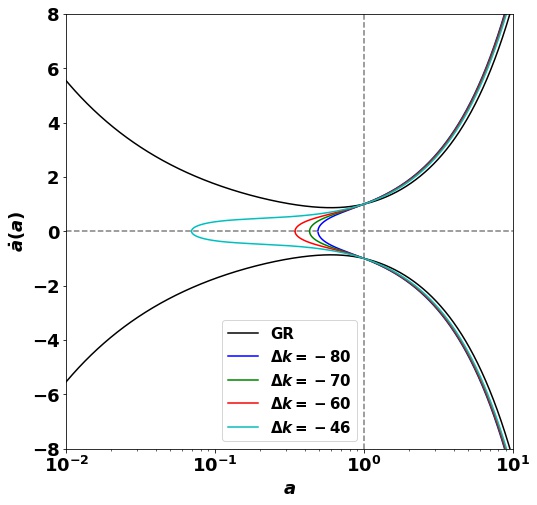}
\caption{\footnotesize The ``phase space'' plot of the expansion rate $\dot{a}$ (in units $H_0$) vs. the scale parameter $a$. The transition from collapse (lower branch with $\dot{a} < 0$) to expansion (upper branch with $\dot{a} > 0$) is smooth implying a hold at a finite size (non-singular, soft bounce).}
 \label{fig:Iphasespace}
 \end{minipage}
 \end{figure}

In this case $K_0 < \bar{K}_0$, and both, $g_2$ and $g_1$, are finite and positive.  
The scale potentials $V(a;g_2(K_0))$, \eref{V(a)}, for the various negative values of the parameter $K_0 < \bar{K}_0$ are plotted in \fref{fig:IV(a)}. 

The potential becomes greater than the total energy $-K_0$ at some turning point $a_i$.  As the dynamics is time-reversal invariant, a possible shrinking of the universe comes to a sudden hold at $a_i$, and is smoothly reversed\footnote{The expansion continues though for ever. The theory as is does not implicate a reversal (at time $t=\infty$?) of that expansion} to an expansion, resulting in a non-singular bounce. That dynamics is summarized in the \emph{phase-space} plot,~\fref{fig:Iphasespace}. Notice that with $K_0 \rightarrow \bar{K}_0$ the turning point $a_i$ moves to smaller scales and larger redshifts $z$. This corresponds to $g_1 \rightarrow \infty$, i.e. to an increasing impact of quadratic gravity. 

 \medskip
Unlike in Einstein-Friedman cosmology, the space-time in CCGG acquires kinetic energy when deformed from the de Sitter\footnote{Its Ricci scalar is positive, $g_2>0$.} ground state. The contribution of that kinetic energy to the overall energy balance increases with increasing deformation of space-time, and also with increasing ``inertia of space-time'' $g_1$.  
Due to the requirement of covariant conservation of the energy-momentum tensor of ordinary matter it is stored in the form of contortion density, aka dark energy, and in the curvature correction that in the Einstein view of the CCGG theory contributes to the (geometric) stress-energy tensor. These are the energy sources driving the inflation. 

\medskip
The conventional scale potential of the flat Einstein-Friedman cosmology (labeled GR), with just the standard (dark) matter and radiation terms and a constant cosmological constant, does not lead to inflation in the early universe. On the other hand, the scale potentials derived from CCGG display both, inflationary behavior and graceful exit to the dark energy era of GR in the late epoch ($a \gg 1$). The graceful exit occurs as the corrections pertinent to the CCGG theory become constant, asymptotically converging to the GR values. (For details of the various contributions to the Hubble function and of the energy conditions see Appendix~\ref{App:K0<0}. Since $\omega < -1$ in the early epoch, all energy conditions are violated.) 

\begin{figure}[H]
\begin{minipage}{14pc}\vspace{0.1pc}
\includegraphics[width=14pc]{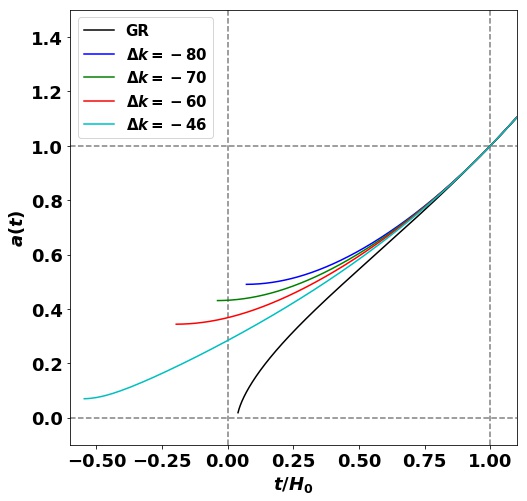}
 \caption{\footnotesize 
 The lin-lin plot of the expansion scale as a function of the universal time. It is finite at the time of birth of the universe that occurs at~$a_i \equiv a(t_i) > 0$ and~$\dot{a}(t_i) = 0$. GR refers to the Einstein-Friedman cosmology.}
 \label{fig:I_a(t)}
 \end{minipage}\hspace{1pc}%
 \begin{minipage}{14pc}
\includegraphics[width=14pc]{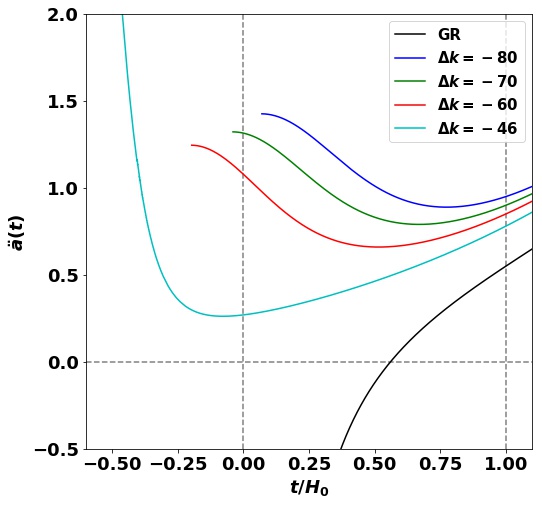}
\caption{\footnotesize The acceleration $\ddot{a}(t)$ (units $H_0^2$) as a function of the universal time $t$. It is always positive indicating an ever lasting inflation of the universe. For $K_0 \gtrsim -70$ the age of the universe exceeds the age inferred from Einstein gravity.}
 \label{fig:I_add(t)}
 \end{minipage}
 \end{figure}

  \medskip
  The scale parameter as a function of the universal time, $a(t)$, is given by the integral  
\begin{equation}
 t - t_i  = \int_{a_i}^{a} \left[ -K_0 - V(a';f(a';K_0)) \right]^{-1/2} \, da',
\end{equation}
derived from~\eref{DG} where the solution~$f(a;K_0)$ of~\eref{ODE:f(a)} is inserted. We fix the global time scale by setting $t(1) = 1/H_0$. Then $\dot{a}(t)$ is  given by inserting $a(t)$ (cf.~\fref{fig:I_a(t)})  into $\sqrt{-K_0-V(a)}$. While the so calculated expansion rate at time $t_i$  vanishes, i.e.~$\dot{a}(t_i) = 0$, the acceleration  is positive there, $\ddot{a}(t_i) > 0$, as shown in the plot~\fref{fig:I_add(t)}.  The universe thus evolves from a non-singular, soft bounce (``Little Bang'') through an ever lasting inflation, with phases of varying intensity, and finally, via a graceful exit into the dark energy era. The initial scale $a_i$ of the universe decreases with $g_2 \rightarrow 0$, while its age and the violence of the explosion, $\ddot{a}_i$, increase. The "soft" characteristic, namely $\dot{a}(t_i) = 0$, of this non-singular bounce scenario is maintained, though.

\medskip
The scenario thus encountered in the parameter Region I has the following characteristic:
\begin{enumerate}
 \item No singularity - the universe starts off from a finite volume,
 \item Soft bounce with initial zero expansion rate but non-zero acceleration, 
 \item Steady inflation commencing after the bounce, 
 \item ``Graceful exit'' to the late dark energy era, similar to Einstein gravity.
\end{enumerate}
We call it the \emph{Non-singular Bounce} scenario.

\subsection{Region II: Hard singular Big Bang with a secondary inflation-deceleration phase} \label{sec:RegionII}
\begin{figure}[H]
\begin{minipage}{14pc}
 \includegraphics[width=14.5pc]{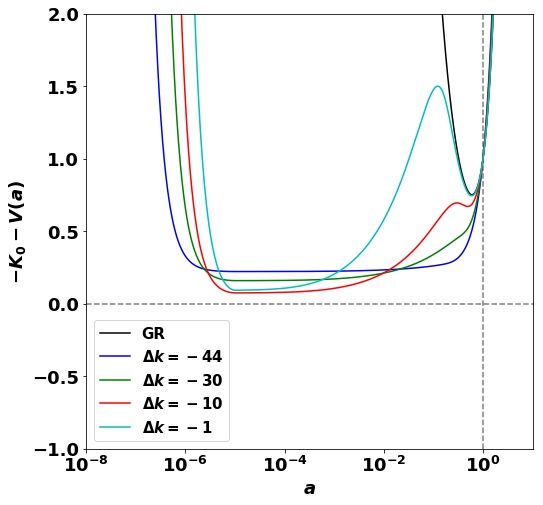}
 \caption{\footnotesize 
 The scale potentials $-K_0 - V(a;g_2(K_0))$ in units of $H_0^2$. Since $-K_0-V(a;g_2(K_0)) > 0$ always holds, there is no turning point.
 The curve labeled GR is the scale potential of the conventional flat Einstein-Friedman cosmology ($f(a) \equiv 1, g_1 = 0$). }
 \label{fig:IIV(a)}
 \end{minipage}\hspace{1pc}%
 \begin{minipage}{14pc}
 \includegraphics[width=14pc]{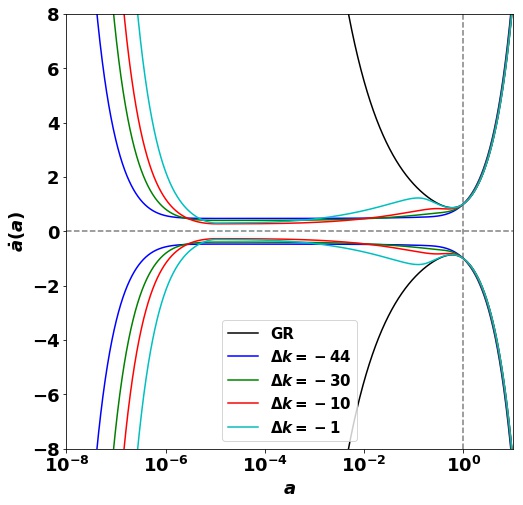}
\caption{\footnotesize The ``phase space'' plot of the expansion rate $\dot{a}$ (in units $H_0$) vs. the scale parameter $a$. The expansion phase with $\dot{a} >0$ is displayed in the upper panel, while the lower panel corresponds to a collapse towards $a=0$.}
 \label{fig:IIphasespace}
 \end{minipage}
 \end{figure}

In this case $0 > K_0 > \bar{K}_0$, and both $g_2$ and $g_1$ are finite and negative (dS~geometry), i.e. the kinetic energy from  the quadratic term in the Hamiltonian has the opposite sign compared to Region I. With $K_0 \rightarrow \bar{K}_0$ the limiting values for the coupling constant is $g_2 \rightarrow 0$, i.e. equivalently $g_1 \rightarrow -\infty$ which means a dominant quadratic gravity.  The scale potentials, plotted as $-K_0-V(a;g_2(K_0))$ in \fref{fig:IIV(a)}, differ from the GR scale potential but never exceed the ``total energy'' $-K_0$.

\begin{figure}[H]
 \begin{minipage}{14pc}
\includegraphics[width=14pc]{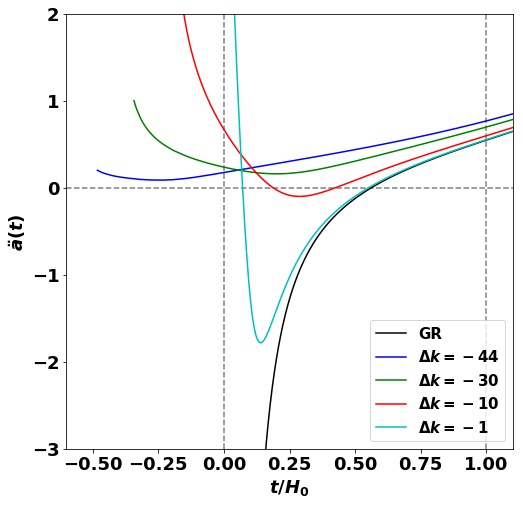}
\caption{\footnotesize The acceleration $\ddot{a}(t)$ (units $H_0^2$) as a function of the universal time $t$. The acceleration of the inflation phase goes over into a deceleration period with $\ddot{a}(t) < 0$, and rebounds again at a lower level in the late era.}
 \label{fig:II_add(t)}
 \end{minipage}\hspace{1pc} 
\begin{minipage}{14pc}\vspace{.05pc}
\includegraphics[width=14.7pc]{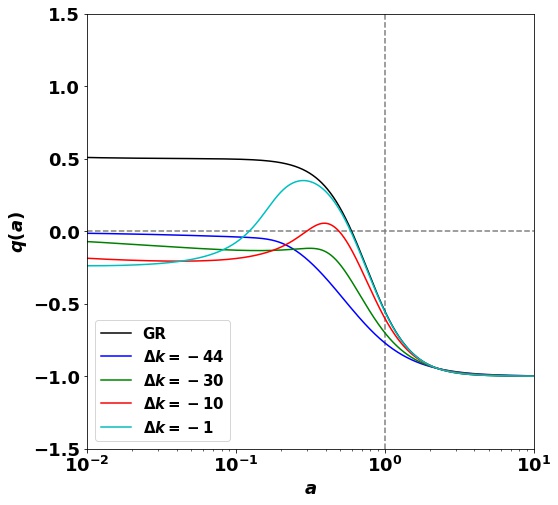}
 \caption{\footnotesize The log-lin plot of the dimensionless deceleration parameter~$q(a;K_0<0)$ showing the region of the most recent transition from deceleration to acceleration of the expansion of the universe. The transition scale $a^*$ is defined by $q(a^*) = 0$. 
 }
 \label{fig:II_q(a)}
  \end{minipage}

 \end{figure}

\begin{figure}[H]
\begin{minipage}{14pc}\vspace{0.8pc}
\includegraphics[width=14pc]{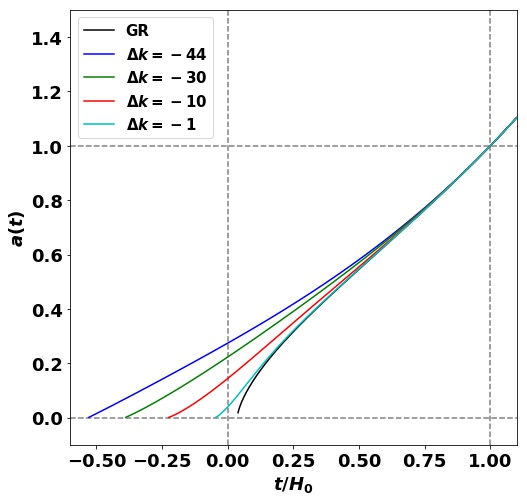}
 \caption{\footnotesize 
 The lin-lin plot of the scale parameter as a function of the universal time indicating a singular Big Bang and a finite age of the universe. GR refers to Einstein gravity.}
 \label{fig:II_a(t)}
  \end{minipage}\hspace{1pc}%
 \begin{minipage}{14.7pc}
\includegraphics[width=14.3pc]{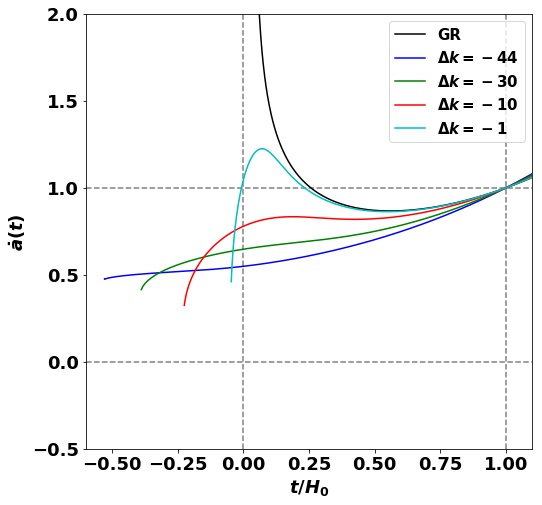}
 \caption{\footnotesize The expansion rate $\dot{a}(t)$ (units $H_0$) commences with a non-zero value and diminishes as a function of the universal time.}  
 \label{fig:II_ad(t)} 
 \end{minipage}
\end{figure}

\medskip
Moreover, the asymptotic behavior, $V(a \rightarrow 0) = -\infty$, is similar to Einstein cosmology, and we expect this type of universe to be singular. Indeed, the phase-space plot~\fref{fig:IIphasespace} confirms this expectation. Interpreting the phase-space diagram as if the birth of the universe had occurred after a Big Crunch, it resembles a supernova explosion where the kinetic energy of the collapsing matter drives the subsequent explosion. 
%
Unlike the non-singular Bounce scenario with negative $K_0$ and positive $g_2$ discussed in \sref{sec:RegionI}, the initial acceleration at time $t_i$ is negative. It changes sign though at $a ~\sim 10^{-3}$. The following inflation-deceleration phase is depicted in~\fref{fig:II_add(t)}. The initial expansion thus slows down and terminates at the first peak of the potential, with a secondary inflation that terminates when the scale hits the bottom of the potential well \cite{benisty19d}, after which a second deceleration phase commences. 
That deceleration stops at the top of the second peak of the scale potential, followed by soft re-acceleration that reaches the value $q_0 = -0.5 + K_0/H_0^2$ today and continues into the dark energy era. 
The corresponding deceleration parameter $q(a)$, see~\fref{fig:II_q(a)}, illustrates the acceleration-deceleration transition in more detail. Its present-day value decreases linearly with $K_0$.
In Ref.~\cite{riess04} the most recent deceleration-acceleration transition has been identified at redshift in the region~$z^* \approx 0.3 - 0.7$, which corresponds to the scale $a^* = 1/(1+z^*) \approx 0.6 - 0.8$.  The authors of Ref.~\cite{riess04} also conjecture that the cosmological constant must be time dependent. 
Notice that the local extrema of the potential become more pronounced and shift to smaller scales and earlier times with $K_0 \rightarrow 0_-$. 
The singularity of the Big Bang is illustrated in the plot~\fref{fig:II_a(t)} where the scale parameter is shown to asymptotically reach zero at some finite time $t_i(K_0)$, in general exceeding $H_0^{-1}$. The ``hard'' character of the Big Bang is attributed to the initial non-zero expansion rate, cf.~\fref{fig:II_ad(t)}.   

A \emph{Singular ``Hard'' Big Bang} scenario thus emerges in the parameter Region II. It is initially similar to Einstein-Friedman cosmology but with a ``built-in'' inflation phase. It has the following characteristic: 
\begin{enumerate}
 \item Singularity $a(t_i) = 0$,
 \item Violent initial expansion with diminishing acceleration, similar to conventional Big Bang, and comparable to a supernova explosion,
 \item A secondary inflation-deceleration phase, 
 \item Graceful exit into a second late inflation epoch (dark energy era) at $a^* \approx 0.6$. 
 \end{enumerate} 

 \medskip
 The overall behavior of the scale expansion is again driven by the interplay of the correction caused by the dynamics of the space-time geometry leading to a negative effective pressure, see Appendix~\ref{App:K0<0}. Notice that all energy conditions are satisfied except SEC.

\subsection{Region III: Singular hard Big Bang without inflation}
\begin{figure}[H]
\begin{minipage}{14pc}\vspace{-.1pc}
\includegraphics[width=14pc]{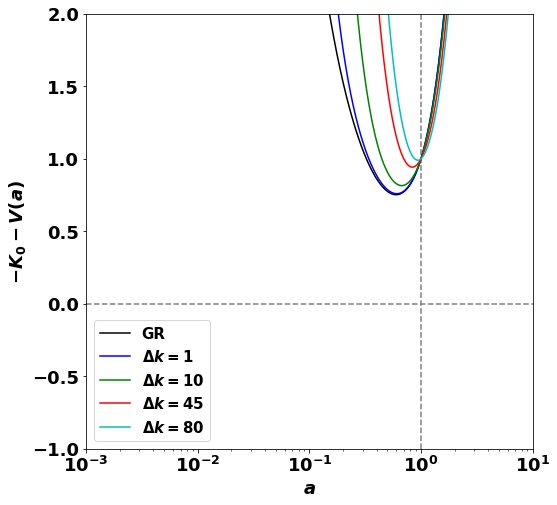}
 \caption{\footnotesize 
 The log-lin plot of the scale potentials $-K_0-V(a;g_2(K_0))$ in units of $H_0^2$.
 The curve labeled GR shows the potential of the conventional flat Einstein-Friedman theory ($f(a) \equiv 1, g_1 = 0$). }
 \label{fig:IIIV(a)}
 \end{minipage}\hspace{1pc}%
 \begin{minipage}{14pc}\vspace{-1.75pc}
\includegraphics[width=14pc]{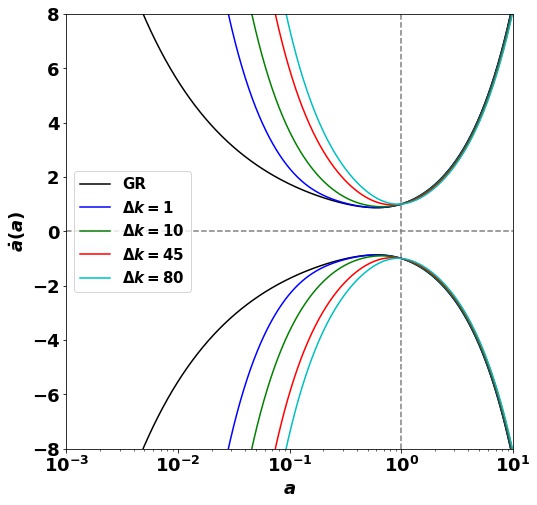}
\caption{\footnotesize The ``phase space'' plot of the expansion rate $\dot{a}$ (units $H_0^2$) vs. the scale parameter $a$. }
 \label{fig:IIIphasespace}
 \end{minipage}
 \end{figure}

\begin{figure}[H]
\begin{minipage}{14pc} 
\includegraphics[width=14.3pc]{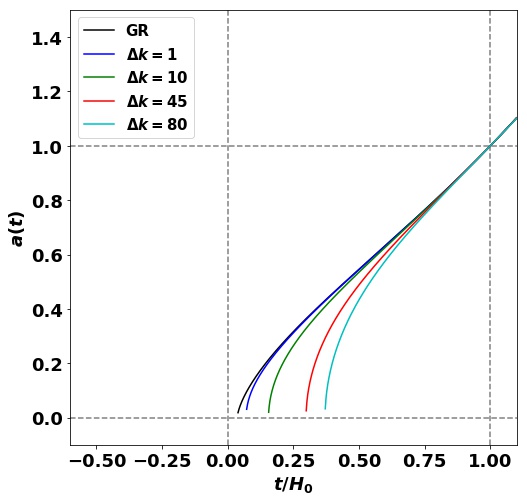}
 \caption{\footnotesize 
 The lin-lin plot of the scale parameter as a function of the universal time indicating a singular Big Bang and a finite age of the universe. GR refers to Einstein-Friedman cosmology.}
 \label{fig:III_a(t)}
  \end{minipage}\hspace{1pc}%
 \begin{minipage}{14pc}
\includegraphics[width=14pc]{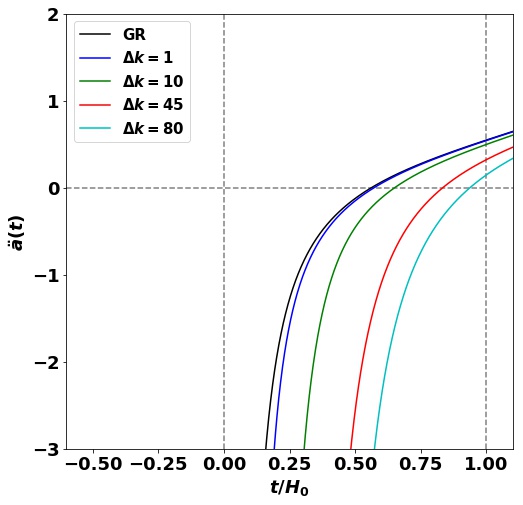}
 \caption{\footnotesize The expansion acceleration $\ddot{a}(t)$ (units $H_0^2$) commences with a large negative value at the initial time $t_i$ but the deceleration decreases with increasing universal time.}  
 \label{fig:III_add(t)} 
 \end{minipage}
\end{figure}

In the parameter Region III, $K_0 > 0$ and both $g_2$ and $g_1$ are finite and positive.
For $K_0 \rightarrow 0$ the limiting value for the coupling constant is $g_2 \rightarrow \infty$ (and equivalently $g_1 \rightarrow 0$).
The resulting scale potentials, displayed in~\fref{fig:IIIV(a)}, give rise to a singular Big Bang scenario as in Einstein gravity. The Einstein-Friedman scenario is asymptotically approached with~{$\Delta k = K_0 \rightarrow 0$} as seen also from the phase-space plot,~\fref{fig:IIIphasespace}.  
The singularity and the violent character of the Big Bang are obvious as $a(t_i) = 0$ and $\ddot{a}(t_i) \ne 0$, cf. the plots~\fref{fig:III_a(t)} and ~\fref{fig:III_add(t)}. The acceleration is negative in the early phase, hence there is no inflation caused by the dynamical space-time geometry. The character of the initial explosion is again supernova-like, but unlike the dynamics in the parameter Region II, no secondary inflation phase arises. This scenario is thus 
a copy of the conventional Einstein-Friedman cosmology but its dynamics including the onset of the dark energy era is shifted to smaller redshifts --- presumably at odds with observations.  
The general behavior of the dark energy and curvature functions,~$f(a;K_0)$ and $k(a;K_0)$, for positive~$K_0$ (and positive~$g_2$) is discussed in Appendix~\ref{App:K0>0}. The energy conditions NEC, WEC and DEC are satisfied. 

\subsection{Comparison with the SNeIa Hubble diagram} \label{sec:SNeIa}
\begin{figure}[H] 
\begin{tabular}{cc}
  \includegraphics[width=14pc]{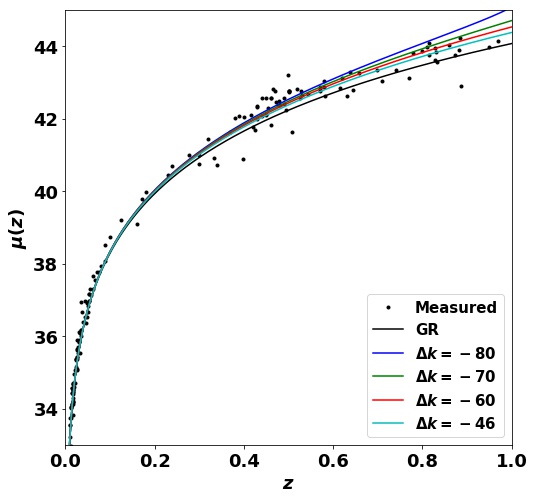} &   \includegraphics[width=14pc]{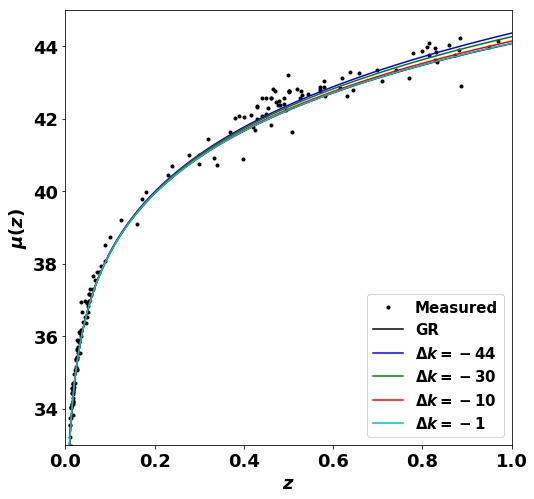} \\
\includegraphics[width=14pc]{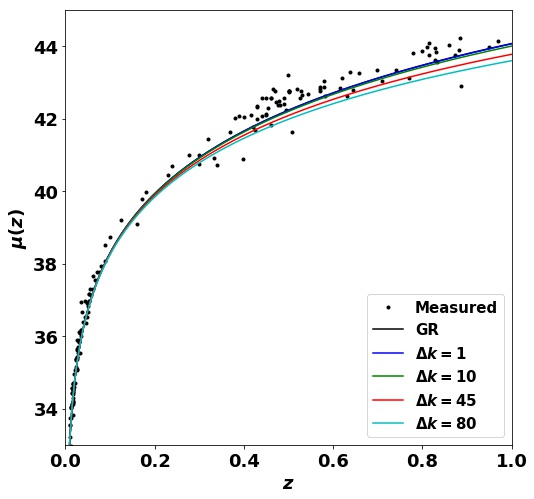} &   \includegraphics[width=14pc]{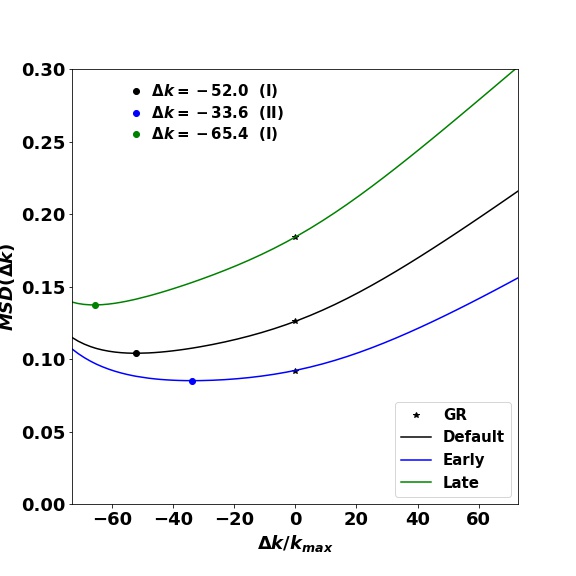} \\
\end{tabular}
\caption{\footnotesize  The SNeIa Hubble diagrams are compared with the model prediction for the Regions I (top left), II (top right), III (bottom left). The plot at the bottom-right is the mean-square deviation for various sets of the $\Lambda$~CDM parameter sets.} 
\label{fig:LD}
\end{figure}

We finally compare the CCGG cosmology model and the standard GR $\Lambda$CDM model with the observational data via the relation between the distance modulus~$\mu$ and the redshift~$z$. The observational supernovae data come from Ref.~\cite{riess04}. Distance estimates from SNeIa light curves are derived from the luminosity distance
\begin{equation}\label{eq:LD1}
 d_{L}=\sqrt{\frac{L_{int}}{4\pi \mathcal{F}}}=a\left(t_0 \right) \,\left(1+z \right)\int_{t_0}^t \frac{dt'}{a\left(t' \right)}
 =\left(1+z \right) \int_0^z \frac{dz'}{H\left(z' \right)},
\end{equation}
where $L_{int}$ is the intrinsic luminosity and $\mathcal{F}$ is the observed flux of the supernovae. 
Inserting the Hubble function for the CCGG model,~\eref{eq:modFriedman4}, into~\eref{eq:LD1} yields
\begin{align}\label{eq:LD3ttt}
&d_{L} = \left(1+z \right)\, \bigintsss_0^z dz' \times \\
& \left[ 
C_m\left(1+z' \right)^{3}
-K_0\left(1+z' \right)^{2}
+C_\Lambda f\left(z' \right)
+\frac{
\left(C_\Lambda f\left(z' \right)
+\frac{1}{4} C_m \left(1+z' \right)^{3}\right)
\left( \frac{3}{4} C_m \left(1+z' \right)^{3}\right)}
{\frac{1}{2} g_{2} -\frac{1}{4} C_m \left(1+z' \right)^{3}-C_{\Lambda}f\left(z' \right)} \right]^{-1/2}.
\nonumber
\end{align}
The logarithm of the luminosity distance is related to the flux (apparent magnitude, $m$) and luminosity (absolute magnitude, $M$) of the observed supernovae via the formula for the extinction-corrected distance modulus, $\mu=m-M=5\log{\frac{d_{L}}{Mpc}}+25$. The dependence of the predicted distance modulus $\mu$ on the redshift $z$ is plotted for the parameter Regions I, II, and III in~\fref{fig:LD}, and compared with the observational data of the SNeIa Hubble diagram~\cite{riess04}.  
Obviously, the CCGG calculations for negative $K_0$ in the vicinity of  $\bar{K}_0/k_{\mathrm{max}} = 45.011$, for which the coupling constant $g_2$ vanishes and $g_1$ is large, are the best fit of the data. The quadratic scale-invariant gravity model thus seems to be preferred over Einstein-Hilbert. The mean-square deviation is minimized for $K_0/k_{\mathrm{max}} = -52$, i.e. in Region I, pointing at a non-singular universe with steady inflation. Yet a sensitivity analysis (bottom right plot) reveals that slight modifications of the concordance parameter set (cf. Table~2) shift the optimum to Region II: For the ``Late'' data sets  we find the MSD minimum at $K_0/k_{\mathrm{max}} = -65.4$, but at $K_0/k_{\mathrm{max}} = -33.6$ for the ``Early'' data set (CMB). Interestingly, the Early data set yields a better fit with the observations of the late epoch\footnote{Whether this affects the Hubble tension remains to be seen. }. Hence the singular Big Bang with a secondary inflation phase is also a good scenario consistent with the Concordance Model, even more as it also satisfies the standard energy conditions.  
 
\begin{table}[ht]
\label{tab:Parameter2}
 \begin{tabular}[t]{|c|c|c|c|c|} 
\toprule
Data & $\Omega_\Lambda$ & $\Omega_m$ & $\Omega_r$ & $h_0$\\
\midrule
Default & $0.69990$ & $0.30000 $ & $0.00005$ & $0.70903$ 
\\
\midrule
\textcolor{black}{Late} & $0.70000$ & $0.30000 $ & $0.00005$ & $0.74500  $
\\
\midrule
Early & $0.68500$ & $0.31500 $ & $0.00005$ & $0.67400  $
\\
\bottomrule
\end{tabular} 
\caption{\footnotesize 
 The $\Lambda$CDM parameter sets used for the sensitivity check of the Hubble diagram fit. The data are taken from the Refs. \cite{planck15} (= Default, applied throughout this paper), \cite{dhawan20} (Late) and \cite{planck18} (Early). The constants used in the equations are given by $C_i = \Omega_i \, h_0^2 \,H_{100}^2$.
 }
 \end{table}
 
\section{The cosmological constant problem} \label{sec:CosmConstProb}

According to Eqs.~\eqref{def:constantsg32} and \eqref{def:cosmfield}, the observable cosmological constant is composed of three \emph{independent} terms: The vacuum energy $g_4/M_p^2$, the ``(A)dS curvature'' $3g_2$, and the present-day value of the Cartan contortion density, $P(1)$. This variety of contributions facilitates sufficient freedom to align the theoretical and observational values of the cosmological constant, and thus provides a new perspective for resolving  the so-called ``Cosmological Constant Problem''\footnote{Similar conclusions with respect to the cosmological constant have been discussed elsewhere, see for example \cite{chen10, ellis13}. However, a thorough cosmological model is missing in that framework.}. Any present-day vacuum energy density $g_4$ can, by a ``suitable'' choice of the \emph{necessarily} non-vanishing (A)dS coupling constant $g_2$ and the Cartan contortion density, be made compatible with the present-day value of the cosmological constant. 

\medskip
To be more specific, provided $g_4 \sim M_p^4 > 0$ and the unknown field $P(x)$ is neglected, the vacuum energy will be compensated to a value close to zero if the deformation parameter is of the order $g_1 \approx 3/2$ \cite{vasak19a}, or $g_2/H_0^2 \approx 10^{120}$. Such a large positive value is achievable with a very small negative value of the FLRW spatial curvature parameter, $K_0 \rightarrow 0_+$. 
Taking though the best fit to the Hubble diagram as discussed in \sref{sec:SNeIa}, the curvature parameter~$K_0 / k_{\mathrm{max}}$ is found in the vicinity of $g_2 \approx 0$. Then the contribution of the Cartan contortion density cannot be neglected in the balance equation \eqref{def:cosmfield}, giving the present-day scalar contortion density of the order $\quarter P(1) \sim M_p^2$. 

\medskip
However, assuming $g_4 = 0$, i.e. abandoning the ``naked'' cosmological constant, 
gives 
for the best-fit parameters $K_0 / k_{\mathrm{max}} = -52.0, -65.4,-33.6$ the respective values 
$\onethird \lambda_0/H_0^2 = g_2/H_0^2 = 0.21, 0.48, -0.62$. This is indeed at the order of magnitude of $\Omega_\Lambda \approx 0.7$, a remarkable fact especially in view of the uncertainty of the $\Lambda$CDM priors and the unknown value of the present-day contortion density. $g_2<0$, pointing to Region~II, is applicable only if $g_4 \le 0$.



\section{Summary and conclusions}

 The CCGG theory facilitates, in a mathematically rigorous way, a consistent description of the dynamics of space-time and matter. Applying the de~Donder-Hamilton-Palatini framework of covariant canonical transformations, it unambiguously fixes the coupling of space-time to matter fields~\cite{struckmeier17a}, and requires a quadratic momentum tensor term~\cite{benisty18a}  extending the Einstein-Hilbert theory. 
 The canonical field equations are obtained by variation of the action integral with respect to the independent fields affine connection, metric and their conjugate momentum fields. Combining these field equations gives the Einstein equation of General Relativity extended by a quadratic Riemann-Cartan concomitant. Since the Schwarzschild and Kerr metrics are solutions of that extended (so-called CCGG) equation, all standard solar tests can be reproduced \cite{struckmeier17a}.
 
\medskip
 Space-time in the CCGG theory is a dynamical medium endowed with kinetic energy and inertia. The strength of that inertia is determined by the dimensionless (deformation) parameter $g_1$ that must be finite in order to ensure the theory's intrinsic consistency.
 The dynamics of matter and space-time are inter-twinned such that only the total energy-momentum, i.e. sum of stress and strain energies, is covariantly conserved. By requiring the stress-energy tensor to satisfy the  covariant conservation law for any given metric, the resulting covariant conservation of the strain-energy tensor is in general possible only with an asymmetric affine connection. That leads to correction terms in the CCGG equation based on the then \emph{necessary} presence of torsion of space-time. In the Friedman universe that torsion-dependent portion of the Ricci curvature scalar (called here Cartan contortion density) emerges as a new dynamical energy reservoir, expressed as a running cosmological constant that we call dark energy field. This  and further curvature corrections invoked by space-time's inertia modify the cosmological dynamics. 
 
 \medskip
 The curvature constant of the FLRW metric, $K_0$, emerges as a new, \textcolor{black}{and the only,} free parameter of the theory\textcolor{black}{, and} determines the deformation parameter $g_1$ and the deceleration parameter $q_0$, thus exposing $g_1$ to direct observations\footnote{\textcolor{black}{For clarity: The deformation parameters $g_1 \sim -1/g_2$ and $K_0$ are equivalent via \eref{prooffne1} reducing the number of new independent parameters to just one.}}.   
 
 \medskip
 Our numerical analysis presented here uses for comparative reasons the $\Lambda$CDM parameter set of the Concordance Model to identify three scenarios for the early evolution of the universe:
 
 Scenario I emerges for negative $K_0$ and negative $g_2$ (and positive $g_1$). It describes a universe that starts off from a non-singular bounce event into a steady inflation phase that gracefully exits into the current dark energy era. 
 An interesting feature of this scenario is, in view of the so called "Hubble tension" \cite{riess19}, the increasing Hubble function.
 
 Scenario II --- for negative $K_0$ and positive $g_2$ --- is a hard singular Big Bang initially similar to Einstein-Friedman cosmology but followed by a secondary inflation that, after a deceleration period, finally exits into the late dark energy era. 
 Both scenarios are consistent with the late epoch SNeIa Hubble diagram. 
 
 This is not the case for Scenario III with positive $K_0$ and negative $g_2$. 
 Here the universe undergoes an evolution that is very similar to the standard Einstein-Friedman cosmology but with the Big Bang shifting to later times with increasing $K_0$. 
 
 \medskip
While for the total effective EOS $\omega \ge -1$ holds in Regions II and III, it is not the case in Region I.  Hence NEC, WEC and DEC are satisfied in Regions II and III but violated in Region I.  
 
\medskip 
 We conclude that the quadratic Riemann-Cartan invariant introduced by the CCGG formalism is a necessary extension of the Einstein-Hilbert theory facilitating a viable contribution to cosmology\footnote{The non-singular solutions of CCGG cosmology has also been investigated in~\cite{benisty18b}.  
 Even if torsion is neglected, the CCGG version of the Dirac equation gives rise to a curvature dependent mass correction that drives inflation~\cite{benisty19c}.  }.  
 \textcolor{black}{The model is, unlike other modified gravity models, \emph{formally derived} from first principles. The dark energy and inflation emerge as effects of the extended geometry, with no matter fields beyond the standard model required. And new light is shed on the cosmological constant problem, and perhaps even on the Hubble tension. } 

 \medskip
  \textcolor{black}{The initial analysis presented here is compatible with the $\Lambda$CDM parameter set, but provides due its additional parameter, more flexibility to reproduce observations. A particular example is the decoupling of the FLRW and apparent curvatures, and the correction of the deceleration parameter. 
 In order substantiate the above conjectures,  
 a comprehensive comparison of the theory with the full body of observational evidence including early-universe data is needed. The impact of contortion-related tensor terms and of spin-carrying matter are further challenging areas for future investigations. } 
 
 \subsection*{Acknowledgments}
This work has been supported by the Walter Greiner Gesellschaft zur Förderung der physikalischen Grundlagenforschung e.V.\, 
DV and JK especially thanks the Fueck-Stiftung for support.  \\
The authors also wish to thank Dirk Kehm for helping with the Hubble diagram analysis, and Horst Stöcker, \textcolor{black}{Andreas Redelbach,} Eduardo Guendelman, David Benisty, Patrick Liebrich, Julia Lienert, Peter Hess and Matthias Hanauske for stimulating discussions. \\
This is a pre-print of an article published in The European Physical Journal Plus. The final authenticated version is available online at: https://doi.org/[DOI: 10.1140/epjp/s13360-020-00415-7].

\pagebreak
\appendix
\counterwithin{figure}{subsection}
\counterwithin{equation}{subsection}
\section*{Appendices}
\addcontentsline{toc}{section}{Appendices}
\renewcommand{\thesubsection}{\Alph{subsection}}

\subsection{Derivation of the Friedman equations} \label{App:FM}

The material content of the Friedman model universe are perfect fluids made of classical particles and radiation. 
The stress-energy tensor for a perfect fluid with the density $\rho$ and pressure $p$ is symmetric. In the comoving frame,
\begin{equation} \label{def:perfectemtensor}
T\indices{^\mu_\nu} = \sum_{i=r,m}\, \mathrm{diag}(\rho_i,-p_i,-p_i,-p_i).
\end{equation}
$\rho_i$ and $p_i$ are functions of the global time $t$ only, and the index $i$ tallies just two basic types of matter, namely particles (``dust'', $i=m$) and radiation ($i=r$). Of course, the particle matter is itself a sum over all standard-model particles.
Radiation, on the other hand, includes not only genuine photon energy density but also contribution from highly relativistic particles where mass is negligible compared to their kinetic energy, i.e. neutrinos.

\medskip
The equation of state (EOS) for a perfect fluid 
is assumed~\cite{weinberg72} to have the generic form
\begin{equation} \label{EOSgeneric}
 p_i = \omega_i\, \rho_i,
\end{equation}
where $\omega_m=0$ and $\omega_r=\onethird$. 
The requirement that the covariant divergence of the energy-momentum tensor of this matter content vanishes  implies the scaling law~\cite{reid02, weinberg72}:
\begin{equation} \label{scalinglaw}
 \rho_i (t) \sim a^{-n_i}  \qquad \Rightarrow \qquad \rho_i\,a^{n_i} = \mathrm{const},
\end{equation}
with the definition $n_i \equiv 3(\omega_i+1)$, i.e. $n_r = 4$  and $n_m = 3$. 

\medskip
We wish now to solve the CCGG-Friedman \eref{eq:modEinstein4}, simplified by setting $P'^{(\mu\nu)} = 0$ and $\xi^{\mu\nu} = 0$, for classical, spinless matter forming perfect fluids:
\begin{equation} \label{eq:modEinstein5}
 g_1\, \bar{Q}^{\mu\nu} 
           + \frac{1}{8\pi G}\,\left[ 
           \bar{G}^{\mu\nu} - g^{\mu\nu} \Lambda(x) \,\right]  
           = \bar{T}^{\mu\nu}.
\end{equation}
The components of the Riemann tensor are derived from the FLRW metric \eqref{def:RWmetric} using the Levi-Civita relation.  The component $\mu = \nu = 0$ of \eref{eq:modEinstein5} yields:
\begin{equation} \label{eq:F2}
 -8\pi G g_1\,\left[\left(\frac{\dot{a}^2+K_0}{a} \right)^2 - 
\ddot{a}^2\right]+\dot{a}^2+K_0-\onethird\Lambda_0\,f(a) a^2
 = \frac{8\pi G}{3} \,a^2\,\sum_{i=r,m}\rho_i.
\end{equation}
Dot denotes the derivative with respect to the universal time $t$. For the $\mu = \nu = 1$ component we obtain in addition
\begin{equation} \label{eq:F3}
 8\pi G g_1\,\left[\left(\frac{\dot{a}^2+K_0}{a} \right)^2 - 
\ddot{a}^2\right]+2a\ddot{a}+\dot{a}^2+K_0-\Lambda_0\,f(a) a^2
 = -8\pi G \,a^2\,\sum_{i=r,m}p_i,
\end{equation}
with $f(a)$ defined in \eref{def:f(a)}. 
For the trace of \eref{eq:modEinstein5}, which is independent of the presence of the traceless quadratic Riemann term, we obtain
\begin{equation} \label{eq:F1}
a\ddot{a} + \dot{a}^2 - 2Ma^2 + K_0 = 0
\end{equation}
where
\begin{align}
 M &:= \onethird\left[2\pi G\,\sum_{i}(\rho_i-3p_i)+\Lambda_0\,f(a)\right] 
 = \onethird\left[2\pi G\,\sum_{i}(\rho_i(4-n_i)+\Lambda_0\,f(a)\right] \label{def:M}
 .
\end{align}
Other combinations of the indices $\mu$ and $\nu$ either vanish or reproduce the above equations.  Moreover, adding \eref{eq:F2} and \eref{eq:F3} yields \eref{eq:F1}, so only two of the three equations need to be considered.

\medskip
To touch base with standard Friedman cosmology we first show that with the quadratic Riemann term omitted (i.e. setting here formally $g_1 = 0$ and $f(a) \equiv 1$, and keeping $\lambda = \Lambda_0$ constant) we recover the conventional Friedman equations. Calculating the difference of the equations (\ref{eq:F2}) -  (\ref{eq:F3}) yields
\begin{equation} \label{eq:F4}
 a\,\Lambda_0  - 3\ddot{a} = 4\pi Ga \sum_{i}(\rho_i+3p_i)
\end{equation}
which can be resolved for $\ddot{a}$:
\begin{equation} \label{eq:secondFriedman}
  \frac{\ddot{a}}{a} = - \frac{4\pi G}{3}  \sum_{i}(\rho_i+3p_i) + \onethird\,\Lambda_0.
\end{equation}
In a similar way we obtain from \eref{eq:F1} with the definition of the Hubble function $H := \dot{a}/a$:
\begin{equation*}
  \frac{\ddot{a}}{a} + H^2 + \frac{K_0}{a^2}= \twothird \left[ 2\pi G \sum_{i}(\rho_i-3p_i) + \Lambda_0\right].
\end{equation*}
Inserting \eref{eq:secondFriedman} yields the well known Friedman equation based on General Relativity:
\begin{equation} \label{eq:firstFriedmana}
  H^2(a) = \frac{8\pi G}{3}  \sum_{i}\rho_i - \frac{K_0}{a^2} + \onethird\Lambda_0.
\end{equation}

 It is rather straightforward to re-run the derivation 
 with $g_1 \ne 0$ and look for modifications arising from the quadratic term in the CCGG equation. 
Inserting \eref{eq:F1}  in \eref{eq:F2} yields, after multiplication with $a^{-2}$, the modified Friedman equation
\begin{align} \label{eq:modFriedman}
  H^2(a) =& - \frac{K_0}{a^2} + \onethird\Lambda(a) + \\
  & +\left[
  \frac{8\pi G}{3}  \sum_{i}\rho_i - \frac{(8\pi G)^2}{3} \,g_1\,M(a)  
\sum_{i}(\rho_i-3p_i)\right]
  \left[1-32\pi G g_1 M(a)\right]^{-1}. \nonumber
\end{align}

\medskip
With the equation of state (\ref{EOSgeneric}) and the scaling properties \eqref{scalinglaw} this equation reads:
\begin{equation} 
  H^2(a) = \frac{8\pi G}{3}  \sum_{i}\rho_i - \frac{K_0}{a^2} + \onethird\Lambda(a)
  +  g_1 \frac{8\pi G}{3}  \sum_{i}\rho_i\,n_i
   \frac{8\pi GM(a)}
   {\left(1-32\pi G g_1 M(a)\right)}.   \label{eq:modFriedman2} 
  \end{equation}
  Obviously, while the first three terms on the r.h.s. formally correspond to the conventional Friedman \eref{eq:secondFriedman}, the last term is generated by the quadratic term in \eref{eq:modEinstein} and vanishes with $g_1 = 0$. 
  
\medskip  
  It is convenient to define the constants
\begin{equation} \label{scalinglaw1}
 C_i := \frac{8\pi G}{3}\,\rho_i\,a^{n_i} = \mathrm{const.}, \qquad i=r, m.
\end{equation}
and
\begin{subequations} \label{scalinglaw2}
\begin{alignat}{4}
   C_K &:= -K_0 \qquad \,\omega_K = -\onethird, \,n_K = 2   \label{def:Ck}\\
   C_\Lambda &:= \onethird\Lambda_0\  \qquad \omega_\Lambda = -1, \,n_\Lambda = 0. \label{def:CLambda}
\end{alignat}
\end{subequations}
The modified Friedman equation (\ref{eq:modFriedman2}) then becomes
\begin{equation} 
  H^2(a) = \sum_{i=r,m,K}\,C_i\,a^{-n_i} + C_\Lambda\,f(a)
  +  g_1\,\frac{32\pi G M(a)}
   {\left(1-32\pi G g_1 M(a)\right)}\, \sum_{i=r,m}\,\quarter C_i\,n_i\,a^{-n_i}. \label{eq:modFriedman3}
\end{equation}
with $M(a)$, \eref{def:M}, re-written as
\begin{equation}
   M(a) = \quarter C_m\,a^{-3} + C_\Lambda\,f(a). \label{def:M1}
\end{equation}

\medskip
The present-day energy densities of matter including dark matter, of radiation including photons and neutrinos, are expressed in the text by the cosmological constant $C_i = \Omega_i \, H_0^2$. Where not stated otherwise we use the "Default" values $\Omega_i$ and $H_0$ listed in  (``Concordance Model`` or ``$\Lambda$ Cold Dark Matter'', $\Lambda$CDM model of standard cosmology, see for example~\cite{planck15,bernal17}), and
\begin{subequations} \label{def:constants}
\begin{alignat}{4}
M_p &= \frac{1}{\sqrt{8\pi G}} \approx 2.428 \times 10^{18}\, \mathrm{GeV} \\
|k_{\mathrm{obs}}| &\le k_{\mathrm{max}} 
:= 0.005 \times H_0^2 
. \label{def:kexp}
\end{alignat}
\end{subequations}
With the initial conditions, $f(1) = 1$ and $k(1) = 0$, the value $H_0 \equiv H(a=1)$
 of the Hubble constant is reproduced from~\eref{eq:modFriedman3}:
\begin{subequations}
\begin{alignat}{4}
 H_0 \approx& 1.499 \times 10^{-42} \mathrm{GeV} = 70.259 \mathrm{\frac{km}{s \,Mpc}} = h_0 \, H_{100} = 0.70259 \,H_{100} \\
H_{100} :=& 100 \mathrm{\frac{km}{s \, Mpc}} = 2.135 \times 10^{-42}\,\mathrm{GeV}.
\end{alignat}
\end{subequations}
\subsection{Derivation of the ODE for $f(a)$} \label{App:f(a)}

In this section we pursue the above analysis of the CCGG-Friedman model, \eref{eq:modFriedman3} or \eref{eq:modFriedman4}, to derive the equation of motion for the dark energy function~$C_\Lambda \,f(a)$.


\medskip
It is helpful at this point to set 
\begin{equation} \label{def:-V(a)}
-V(a) := a^2\,H^2(a) + K_0 
\end{equation}
 and re-write \eref{eq:modFriedman4} in the form of the ordinary, one-dimensional differential equation
\begin{equation} \label{DG}
 \dot{a}^2 + V(a) = -K_0.
\end{equation}
This equation describes the radial expansion dynamics of the universe. Formally, though, also the conservation of the "total energy", $-K_0$, of a fictitious point particle with ``kinetic energy'' $\dot{a}^2$, mass 2,  and ``potential energy''
\begin{equation} \label{V(a)}
 V(a) = -C_m\,a^{-1} - C_r\,a^{-2} -C_\Lambda\,a^{2}\,f(a)
 - `
 \frac{M(a)}{\onehalf g_2-M(a)}\,\left( \threequarter C_m\,a^{-1}+ C_r\,a^{-2}\right).
\end{equation}
 Obviously, any radial motion comes to a hold once $\dot{a} = 0$. Hence  
\begin{equation} \label{eq:forbarea}
V(a) > -K_0
\end{equation}
defines a turning point for the expansion (or contraction) of the universe as the expansion velocity, $\dot{a} = \sqrt{-V(a)-K_0}$, vanishes or even becomes imaginary there. For $\ddot{a} > 0$, i.e. if
\begin{equation}
\frac{V(a)}{a} + 2Ma > 0,
\end{equation}
the radial expansion accelerates. The time period in which this is the case is called inflation. 

\medskip
Differentiating now \eref{DG} with respect to the universal time $t$ yields
\begin{equation} \label{eq:ddot1}
 \ddot{a} = -\onehalf \frac{dV(a)}{da}.
\end{equation}
On the other hand, combining Eqs. \eqref{eq:F1} and \eqref{def:-V(a)} gives 
\begin{equation} \label{eq:ddot2}
 \ddot{a} = \frac{V(a)}{a} + 2Ma.
\end{equation}
Comparing the expressions (\ref{eq:ddot1}) and (\ref{eq:ddot2}) for $\ddot{a}$ we conclude that the identity
\begin{equation} \label{def:identityV(a)}
\frac{dV(a)}{da} + \frac{2}{a} V(a) = -4Ma 
\end{equation}
must hold.
Splitting \eqref{V(a)},
\begin{equation}
 V(a;f;g_1) = V_0(a;f) + V_c(a),
\end{equation}
separates the Einstein-Friedman portion,
\begin{equation}
 V_0(a;f) := -C_m\,a^{-1} - C_r\,a^{-2} -C_\Lambda\,a^{2}\,f(a),
\end{equation}
from the ``curvature correction term'',
\begin{equation}
V_c(a) :=   \frac{M(a;f)}{\onehalf g_2+M(a;f)}\,\left( \threequarter C_m\,a^{-1}+ C_r\,a^{-2}\right),
\end{equation}
that depends on $g_2$ and $f(a)$, and vanishes for $g_1 = 0$, i.e. for $g_2 = \infty$. The identity \eref{def:identityV(a)} applied to $V_0$ gives
\begin{equation} \label{def:identityV_01}
\frac{dV_0(a;f)}{da} + \frac{2}{a} V_0(a;f) = -4aM(a;f) -C_\Lambda \,f'(a)\,a^2,
\end{equation}
where $f' = df/da$. Obviously, \eref{def:identityV(a)} holds for $g_1=0$ and $f(a) = const$., i.e. for the conventional Einstein-Friedman scale potential:  
\begin{equation*} 
\frac{dV_0(a;1)}{da} + \frac{2}{a} V_0(a;1) + 4aM(a;1) \equiv 0.
\end{equation*}
A constant cosmological constant violates the CCGG equations, though, as for the ``curvature correction'' term (and $f(a) = const$, $C_m \ne 0$) we get
\begin{equation*}
\frac{dV_c(a)}{da} + \frac{2}{a} V_c(a) + 4Ma \ne 0.
\end{equation*}
For a non-constant $f(a)$ a straightforward calculation gives
\begin{align} \label{def:identityV_c1}
\frac{dV_c(a)}{da} + \frac{2}{a} V_c(a) = 
&- \frac{\threequarter C_m a^{-2}\,M}{\onehalf g_2 - M}   \\
&+ \frac{\onehalf g_2}{(\onehalf g_2 - M)^2} 
\threequarter C_m a^{-4} \left(\threequarter C_m a^{-1} + C_r a^{-2}\right) \nonumber \\
&- C_\Lambda\,f'(a) 
\frac{\onehalf g_2}{(\onehalf g_2 - M)^2} 
\left( \threequarter C_m a^{-1}+C_r a^{-2}\right). \nonumber 
\end{align}
Inserting  \eref{def:identityV_01} and \eref{def:identityV_c1} into the identity  (\ref{def:identityV(a)}) gives, after some algebra, 
\begin{align} \label{ODE:f(a)1}
 \frac{df}{da} &= 
  \frac{3C_m}{4C_\Lambda} a^{-4} \, \times \\
  &\frac{  
  \onehalf g_2 \left(\threequarter C_m a^{-3} + C_r a^{-4}\right)
  - \left(\onehalf g_2 - \quarter C_m a^{-3} - C_\Lambda \,f \right)
  \left(\quarter C_m a^{-3} + C_\Lambda \,f \right)
    }
  {\onehalf g_2 \left(\threequarter C_m a^{-3} + C_r a^{-4}\right)
    +   \left(\onehalf g_2 - \quarter C_m a^{-3} - C_\Lambda \,f \right)^2   
  }. \nonumber
\end{align}
This is an ordinary first-order differential equation (ODE) that has a unique solution for a given initial condition, here $f(1) = 1$. Notice that this ODE is not changed if we replace the constants $C_i$ by $\Omega_i$, and $g_2$ by $g_2/H_0^2$. 

\subsection{Calculated dark energy and curvature corrections and the resulting EOS}
\subsubsection{$K_0 < 0$} \label{App:K0<0}

\begin{figure}[H] 
\begin{tabular}{cc}
\includegraphics[width=14pc]{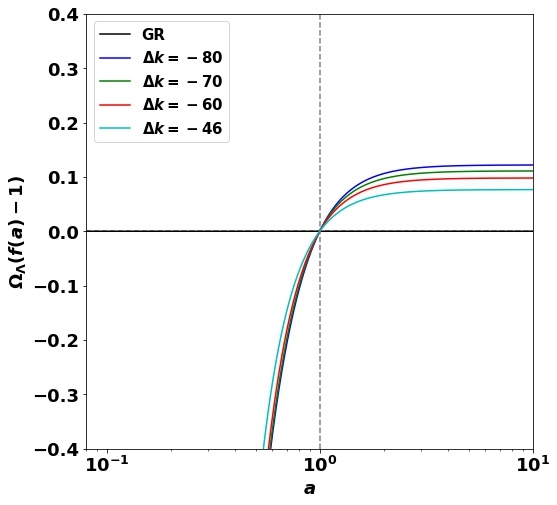} &   \includegraphics[width=14pc]{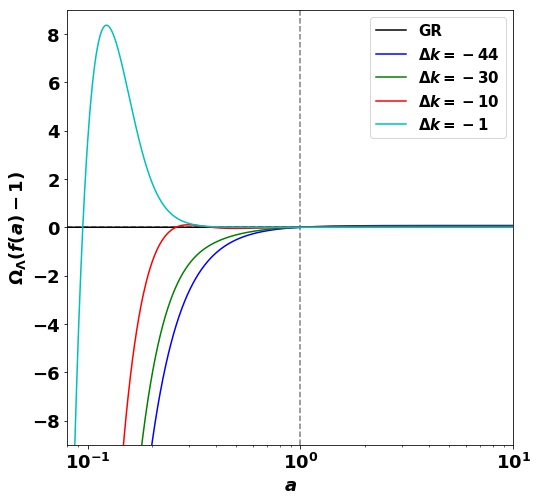} \\
 \includegraphics[width=14pc]{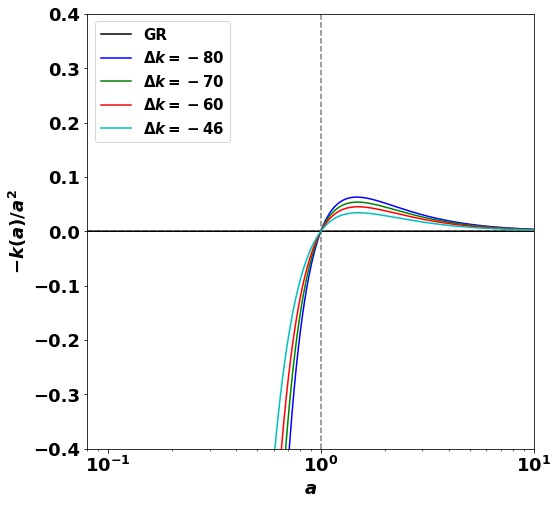} &   \includegraphics[width=14pc]{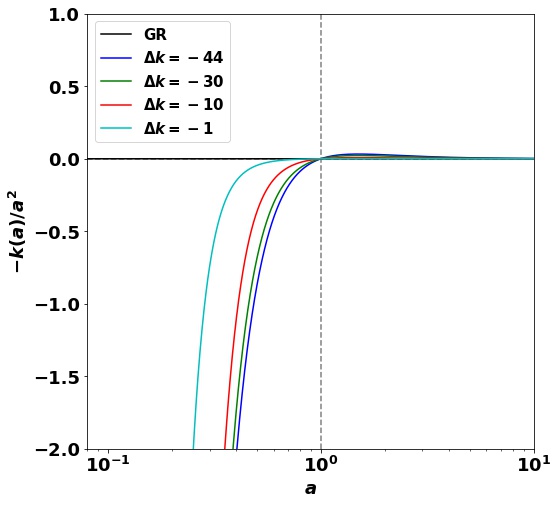} \\
  \includegraphics[width=14pc]{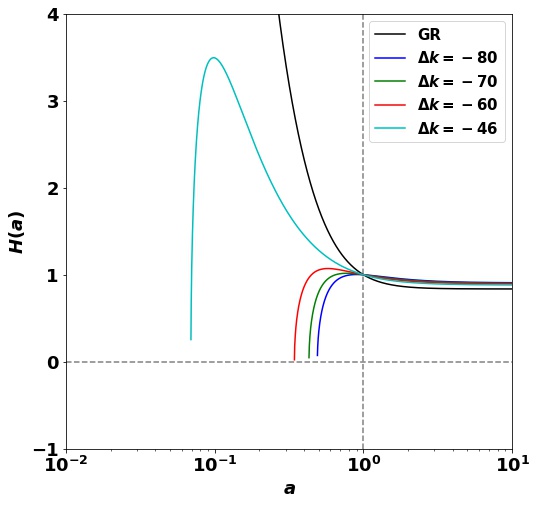} &   \includegraphics[width=14pc]{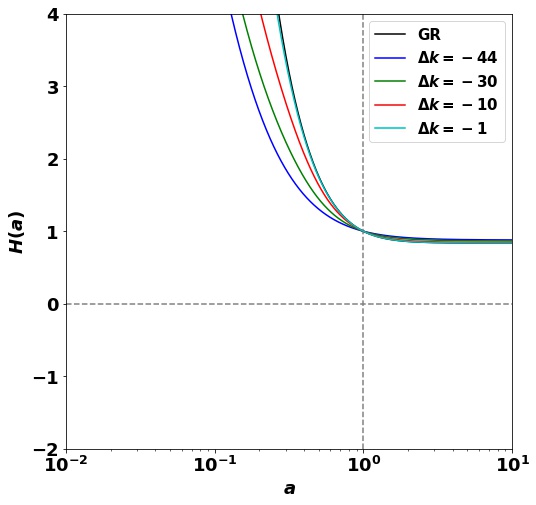} \\
\end{tabular}
\caption{\footnotesize  The scale dependence of the dark energy term (top row), and the curvature  term (middle row) facilitating the geometrical corrections to the Hubble function, and the Hubble function $H(a)$ (bottom row) in parameter Regions I (left column) and II (right column). The units used are 1, $H_0^2$, and $H_0$ respectively.}
\label{fig:fkoI+II}
\end{figure}

The normalized effective dark energy function $f(a;K_0)$ modifies the Hubble function relative to the Einstein-Friedman theory via the correction term $\Omega_\Lambda \,(f(a;K_0)-1)$ plotted in the first row in~\fref{fig:fkoI+II}. It is negative in the early epoch of the universe\footnote{In GR a negative, time-independent cosmological constant in the early universe and cyclic inflation scenarios have been discussed in \cite{biswas09, biswas12}.}. Asymptotically, though, it vanishes paving the way to the dark energy age with $f \sim 1$. This can be understood as an effect of the Cartan contortion density that initially appears to be high and acts as a storage of potential energy. That energy is gradually released to drive the inflation dynamics\footnote{Space-time can be thought of as a knitted elastic medium that can be stretched to some maximum but then rebounds to its equilibrium size. Internally we call our model a universe with ``contorted space-time''.}. 

\medskip
In Region I, with $g_2>0$, the present dark energy reservoir is almost depleted. $f(a)$ reaches its maximum for $a>1$. In Region II, on the other hand, the dark energy function overshoots the asymptotic value of $1$ to large positive values. 
Thus in this case, after complete depletion, the contortion density re-starts absorbing energy again decelerating the expansion until it reaches a saturation value, but with a slightly modified cosmological constant. The overshoot disappears though for $g_2 \rightarrow -\infty$.

\medskip
The effective curvature is, independently of the sign of $g_2$, initially positive, dropping steeply to the present-day value (here $k(1) = k_{\mathrm{obs}} = 0$) to asymptotically arrive at the constant value $K_0$. The plot in the middle row of~\fref{fig:fkoI+II} displays the curvature correction term $-k(a)/(a H_0)^2$ that directly adds to the formula for standard Hubble function $H^2(a)$, and so reinforces the dark energy correction and supports the acceleration of the scale expansion. The corresponding Hubble function is plotted in the bottom row of~\fref{fig:fkoI+II}. The impact of the above correction terms is clearly visible.

\medskip
In Region I the Hubble function $H(a)$ increases from the edge of the forbidden area at $a = a_i$, and passes through a maximum in the recent past\footnote{Interestingly, an increasing Hubble constant is consistent with observational conclusions in Ref.~\cite{risaliti19}.} to approach a constant value $H(a) \approx H_0$ for $a > 1$. For $a \ll 1$ $H(a)$ descents as the correction terms approach $-\infty$ as $a^{-n}$ with the effective power of $n>4$, and thus dominate the radiation term of the Einstein-Friedman cosmology.  
The peak is obviously  the result of a slight mismatch in the $a$-dependence of the dark energy and curvature corrections. 
In Region II the corrections terms diverge with $n \lesssim 4$ and are just a minor correction to the Hubble function.

\begin{figure}[H] 
\begin{tabular}{cc} 
   \includegraphics[width=14pc]{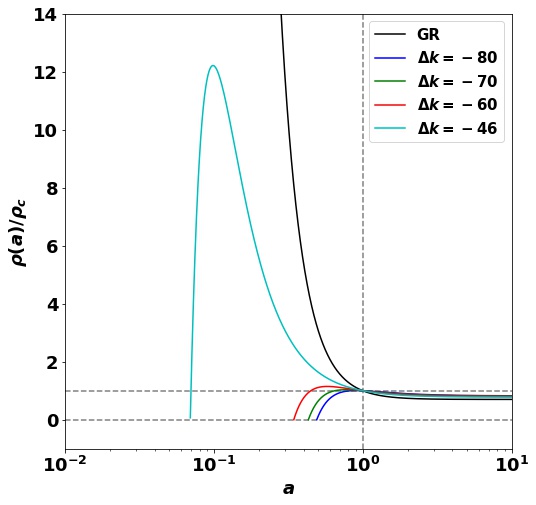} &   \includegraphics[width=14pc]{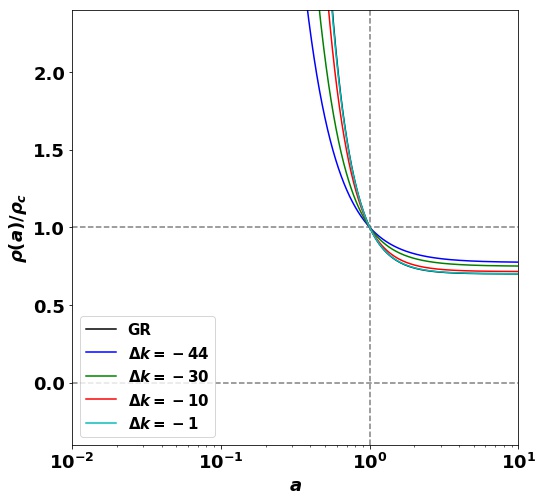} \\
 \includegraphics[width=14pc]{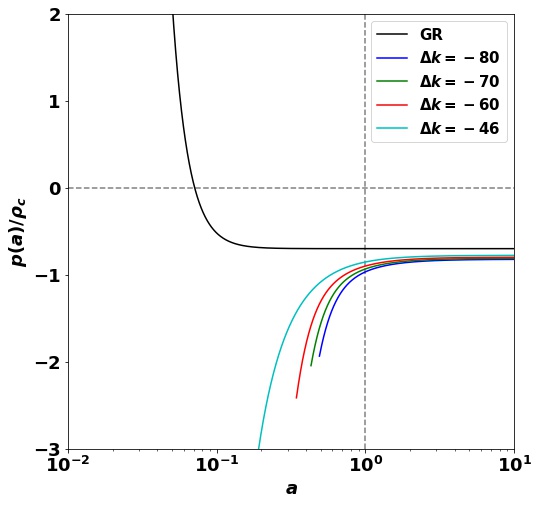} &   \includegraphics[width=14pc]{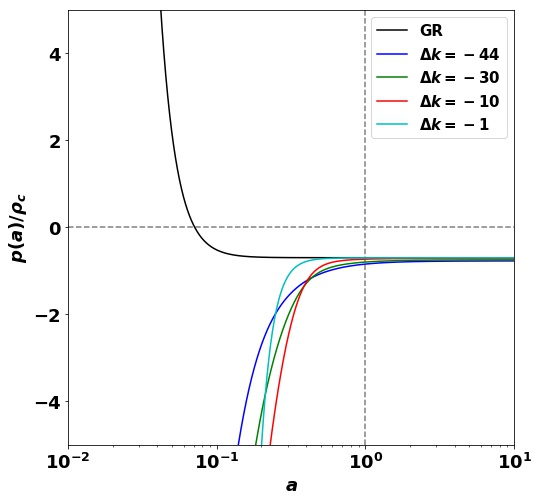} \\
 \includegraphics[width=14pc]{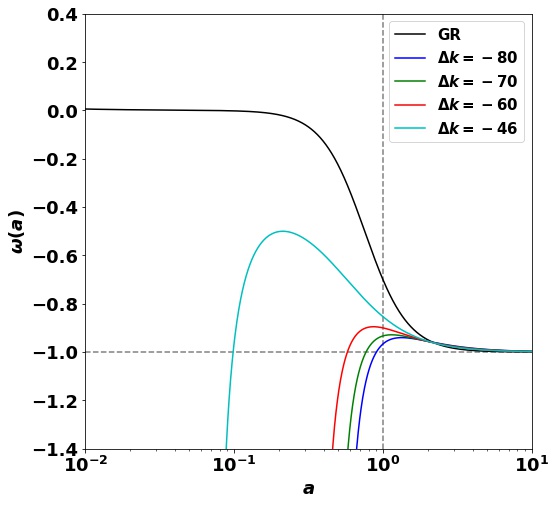} &   \includegraphics[width=14pc]{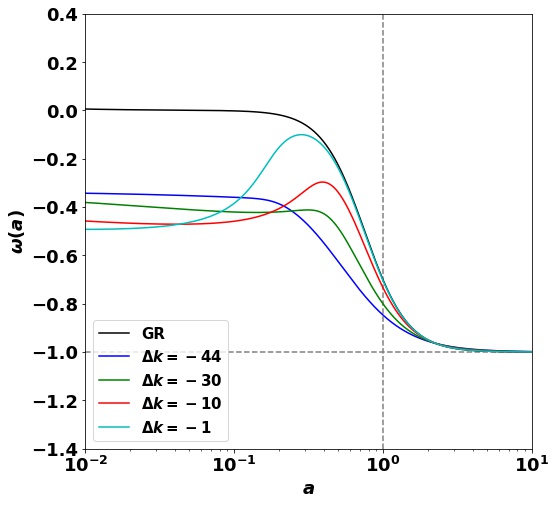} \\
\end{tabular}
\caption{\footnotesize  The scale dependence of the energy density (top row) and pressure (middle row) of the total content of the universum, i.e. (dark) matter, radiation, and geometrical corrections, relative to the critical density $\rho_{crit}$. The bottom row displays the resulting equation of state. Left column: parameter Region I, right column: parameter Region  II. The black lines depict the Einstein-Friedman cosmology. The GR-EOS undergoes a transition from a matter to a dark energy dominated universe. } 
\label{fig:rpoI+II}
\end{figure}

The total energy density of the universe, encompassing matter, radiation and dynamical space-time geometry is always positive, 
see the first row in~\fref{fig:rpoI+II}. The pressure, see the middle row in~\fref{fig:rpoI+II}, is always negative in alignment with ongoing expansion. In Region I the total effective EOS (bottom row) approaches a negative value $\omega(a_i) \ll -1$ at the bounce scale which proves the violation of the standard energy conditions.

\subsubsection{$K_0 > 0$} \label{App:K0>0}
\begin{figure}[H] 
\begin{tabular}{cc} 
   \includegraphics[width=14pc]{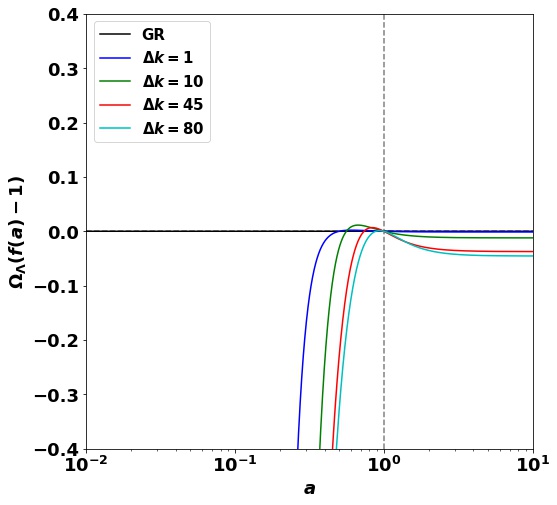} &   \includegraphics[width=14pc]{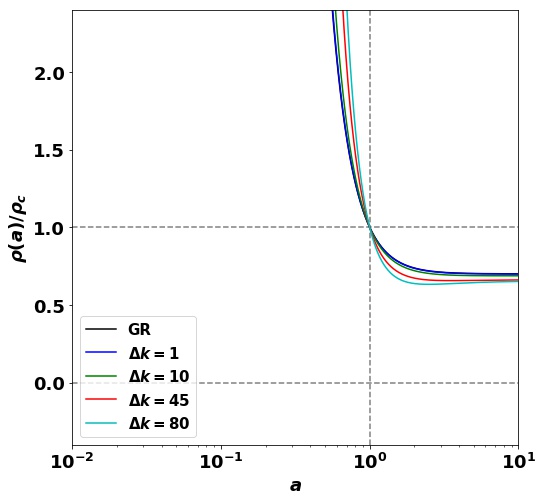} \\
 \includegraphics[width=14pc]{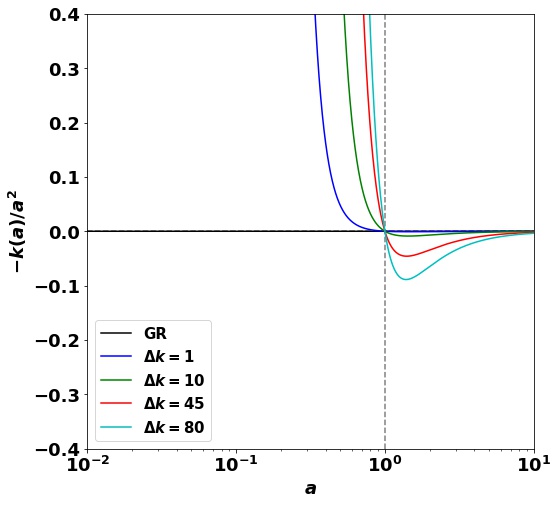} &   \includegraphics[width=14pc]{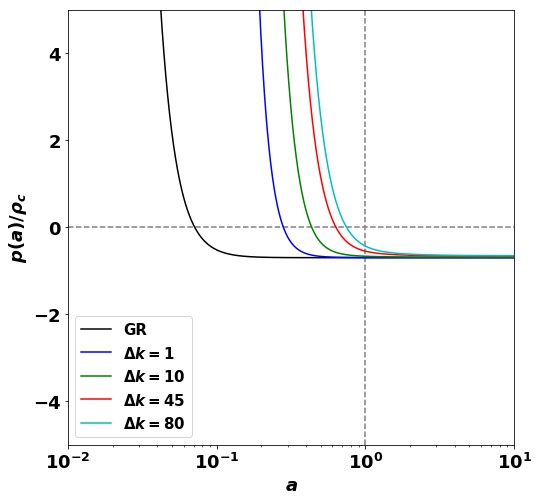} \\
 \includegraphics[width=14pc]{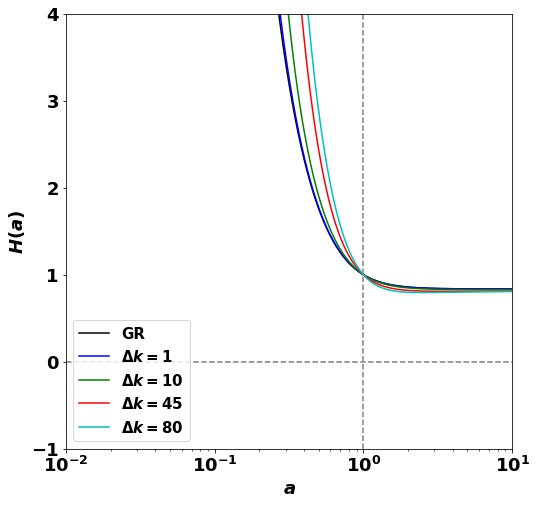} &   \includegraphics[width=14pc]{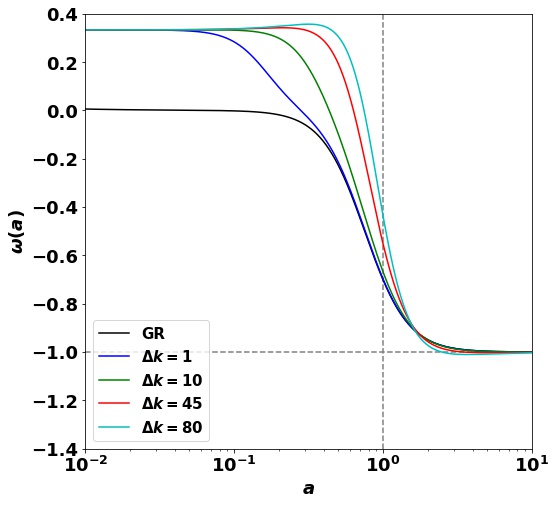} \\
\end{tabular}
\caption{\footnotesize  Parameter Region III. Left column from top to bottom: Geometric corrections of the Hubble function relative to Einstein-Friedman cosmology from the dark energy and curvature functions, and the resulting Hubble function, $H(a)$. The units in the plots are 1, $H_0^2$, and $H_0$ respectively.  Right column: The relative total energy density and pressure, and the resulting equation of state $\omega(a) \ge -1$.} 
\label{fig:rpoIII}
\end{figure}

In Region III the dark energy function and the curvature correction depend on the scale parameter in a similar way, and their contribution to the Hubble function almost cancel out (see the left column of ~\fref{fig:rpoIII}), resulting in an overall evolution of the universe that is rather similar to the flat Einstein-Friedman cosmology. The latter is the limiting case $K_0 \rightarrow \infty$. That similarity is also obvious from the plots of the total energy density, pressure, and EOS (right column). NEC, WEC and DEC but not SEC  are satisfied.

\pagebreak

\end{document}